\documentclass[fleqn,usenatbib]{mnras}

\usepackage{newtxtext,newtxmath}

\usepackage[T1]{fontenc}

\DeclareRobustCommand{\VAN}[3]{#2}
\let\VANthebibliography\thebibliography
\def\thebibliography{\DeclareRobustCommand{\VAN}[3]{##3}\VANthebibliography}

\newcommand{\iboot}{\kappa}

\newcommand{\matr}[1]{\mathbf{#1}}

\newcommand{\nboot}{N_\mathrm{boot}}
\newcommand{\meff}{m_\mathrm{eff}} 
\newcommand{\errmeff}{\Delta m_\mathrm{eff}} 
\newcommand{\dmeff}{d m_\mathrm{eff}}
\newcommand{\errdmeff}{\Delta d m_\mathrm{eff}}
\newcommand{\dmeffi}{d m_\mathrm{eff}^{\iboot}}
\newcommand{\meffi}{m_{\mathrm{eff}}^{\iboot}} 

\newcommand{\ndata}{N_\mathrm{data}}
\newcommand{\var}{\mathit{Var}}
\newcommand{\cov}{\mathit{Cov}}
\newcommand{\ev}{\mathit{E}}

\newcommand{\fsmooth}{\alpha_S}
\newcommand{\dfsmooth}{d\alpha_S}
\newcommand{\ngh}{n_\mathrm{GH}}
\newcommand{\aic}{\mathrm{AIC}}
\newcommand{\aicmod}{\mathrm{AIC_{p}}}

\newcommand{\der}{\mathrm{d}}

\newcommand{\model}{f}

\newcommand{\mcount}{m}

\newcommand{\fitpar}{\model_i}
\newcommand{\chifs}{\chi^2(\fsmooth)}
\newcommand{\dermeff}{\der \meff/\der \fsmooth}
\newcommand{\logdermeff}{\der \meff/\der\log \fsmooth}
\newcommand{\dchii}{d \chi^2_{\iboot}}
\newcommand{\chii}{\chi^2_{\iboot}}

\newcommand{\pfunc}{{\cal P}}

\newcommand{\chipost}{\chi^2_\mathrm{posterior}}
\newcommand{\chiprior}{\chi^2_\mathrm{prior}}
\newcommand{\modelparam}{\Theta}
\newcommand{\bootdat}{z}

\newcommand{\dat}{y}
\newcommand{\dati}{y_i}
\newcommand{\err}{\epsilon}
\newcommand{\erri}{\epsilon_i}

\newcommand{\bfiti}{\model_i(\hat{\modelparam}_\bootdat^{\iboot})}

\newcommand{\bdati}{\bootdat_i^{\iboot}}
\newcommand{\bdat}{\bootdat}

\newcommand{\fiti}{\model_i(\hat{\modelparam}_\dat)}
\newcommand{\fit}{\model(\hat{\modelparam}_\dat)}

\newcommand{\bnoisei}{\Delta z_i^{\iboot}}

\newcommand{\param}{\modelparam}
\newcommand{\fittedparam}{\hat{\modelparam}_\dat}
\newcommand{\fittedparamk}{\hat{\modelparam}_{\dat,k}}
\newcommand{\bootedparam}{\hat{\modelparam}_\bootdat}
\newcommand{\bootedparamiota}{\hat{\modelparam}_\bootdat^\iboot}

\newcommand{\bootedparamkiota}{\hat{\modelparam}_{\bootdat,k}^\iboot}
\newcommand{\bootedparamjiota}{\hat{\modelparam}_{\bootdat,j}^\iboot}

\newcommand{\like}{\cal L}
\newcommand{\logl}{\log {\like}}

\usepackage{graphicx}	
\usepackage{amsmath}	

\title[Smoothing via model selection]{A simple data-driven method to optimise the penalty strengths of penalised models and its application to non-parametric smoothing}

\author[J. Thomas \& M. Lipka]{
Jens Thomas,$^{1,2}$\thanks{E-mail: jthomas@mpe.mpg.de}
Mathias Lipka$^{1,2}$
\\
$^{1}$Max-Planck-Institut für extraterrestrische Physik, Giessenbachstrasse, D-85748 Garching\\
$^{2}$Universitäts-Sternwarte München, Scheinerstrasse 1, D-81679 München, Germany
}

\date{Accepted XXX. Received YYY; in original form ZZZ}

\pubyear{2015}

\begin{document}
\label{firstpage}
\pagerange{\pageref{firstpage}--\pageref{lastpage}}
\maketitle

\begin{abstract}
Information of interest can often only be extracted from data by model fitting. When the functional form of such a model can not be deduced from first principles, one has to make a choice between different possible models. A common approach in such cases is to minimise the information loss in the model by trying to reduce the number of fit variables (or the model flexibility, respectively) as much as possible while still yielding an acceptable fit to the data. Model selection via the Akaike Information Criterion (AIC) provides such an implementation of Occam's razor. 
We argue that the same principles can be applied to optimise the penalty-strength of a penalised maximum-likelihood model. However, while in typical applications AIC is used to choose from a finite, discrete set of maximum-likelihood models the penalty optimisation requires to select out of a continuum of candidate models and these models violate the maximum-likelihood condition. 
We derive a generalised information criterion $\aicmod$ that encompasses this case.
It naturally involves the concept of effective free parameters which is very flexible and can be applied to any model, be it linear or non-linear, parametric or non-parametric, and with or without constraint equations on the parameters. We show that the generalised $\aicmod$ allows an optimisation of any penalty-strength without the need of separate Monte-Carlo simulations. 
As an example application, we discuss the optimisation of the smoothing in non-parametric models which has many applications in astrophysics, like in dynamical modeling, spectral fitting or gravitational lensing.
\end{abstract}




\section{Introduction}
Very often, the information one aims to extract from a set of data points is not an observable itself. Instead, one has to infer the information by fitting a {\it model} to the observed data. Sometimes, when one has a clear understanding of the processes involved in generating the data the functional form of the model can be deduced from first principles. In this case, one is only faced with the problem of finding the optimal parameters of the model whereas the form of the model is fixed. 

In other cases, however, one may not have such a clear picture that allows one to deduce the form of the model. Instead, one may have measured two quantities $x$ and $y$ which happen to be correlated without knowing the underling form of the relation of $x$ and $y$. Then the first question becomes {\it how} this correlation can be characterized: For example, can it be described by a linear model (2 parameters), a parabolic model (3 parameters) or a 3rd-order polynomial (4 parameters), etc.? A simple principle often followed in this task of \textit{model selection} is Occam's Razor, which is to choose the model with the smallest number of free parameters (the "simplest" one) that still describes the data well. Mathematical implementations of this principle are offered by information theory: One of them, the Akaike Information Criterion (AIC, \citealt{Akaike73,Akaike74}) is frequently used to judge models based upon the increase of model complexity (and, hence, of information {\it loss}) against the improvement in the goodness-of-fit (cf. Sec.~\ref{sec:model_selection}).

Many applications of model selection deal with categorically or structurally different models like in the order-selection problem briefly described above. The model selection character of such problems is evident, not least because the models that are compared are represented by {\it different fitting functions}. However, a model is not only characterised by its fitting function. Equally important are the model {\it parameters}. Are the parameter ranges unlimited or subject to equality or inequality constraints? And can the parameters vary independently or are they subject to implicit correlations? The answers to these questions can change the behaviour of a model effectively as much as a change in the fitting function can do.

Consider for example a third order polynomial as briefly mentioned above. It has four model parameters and its model complexity is much higher than that of a straight line with just two parameters. Suppose that for some reason the fit with this 3rd order model is subject to a penalty. A penalty proportional to the square of the second derivative is frequently used in non-parametric fits to keep a model smooth and to prevent it from overfitting the data, i.e. fitting the noise. Minimising the second derivative -- the curvature -- means to make the model a straight line in our case. 
A penalty like this often comes along with one or more additional parameters that allow to adjust the relative strength of the penalty over the achievable goodness-of-fit. In the above example, when we choose the penalty strength such that the weight of the penalty actually vanishes then the model behaves like an ordinary 3rd order fit with four independent parameters. However, if we choose the penalty strength such that it dominates the fit then the same fitting function will effectively behave like a linear model with only two parameters. 

Classically, one considers the penalty as a modification of a {\it specific model} and the penalty strength as a nuisance parameter of that model.  
However, the above illustrates that for a model's behaviour a change of the fitting function or a change of the penalty strength can have equivalent effects. It seems therefore promising to try and view penalised maximum-likelihood models from a slightly different perspective and to reinterpret the penalty as being a function that implicitly spawns a whole new {\it family of different models}. While different fitting functions will typically lead to a discrete set of candidate models, the penalty term allows to generate a continuum of models. The penalty strength is the natural parameter in this continuum to distinguish between different models.
Treating the penalty strength as a parameter of a family of models transforms the task of optimising this strength into a model selection problem. In other words, it allows to adopt model selection techniques to solve the optimisation problem of the penalty strength. The main challenge when dealing with a continuum of models generated via a penalty term as described above is that the fitting function is constant. Thus, the number of fitted parameters is invariable and cannot serve as a measure of model complexity anymore. Hence, one needs to adapt the model selection strategy.

The goal of this paper is to elaborate the above outlined ideas in detail. We show how the classical ideas of model selection can be generalised to models that do not fulfill the maximum-likelihood condition. This includes penalised maximum-likelihood models in particular. Our generalisation can be naturally formulated using an intuitive, generalised concept of free parameters. We show how the generalised model selection can be used to optimise the penalty strength of penalised models. As a specific example application we provide a simple recipe that is based on information theory and that allows to optimise the smoothing of any model. The method is purely based on the measurement data at hand and does not require separate Monte-Carlo simulations (e.g. for calibration of the smoothing). It is flexible and can be applied to linear as well as non-linear models and to models with or without constraint equations on their parameters. We will use the example of an emission line model to illustrate the method and the underlying concepts. It should be noted that the method is however not restricted to smoothing problems and can be applied to any penalised-maximum-likelihood model.

In Sec.~\ref{sec:input} we introduce the mock data set that is inspired by the problem of fitting the shape of an emission line. In Sec.~\ref{sec:recovery} we introduce two example models intended to describe these mock data: a non-linear parametric model and a linear non-parametric model. In Sec.~\ref{sec:model_selection} we recall the basics of model selection for maximum-likelihood models without penalties. In Sec.~\ref{sec:meff} we introduce a bootstrap method to compute the effective number of parameters that can be applied to a large class of models. In Sec.~\ref{sec:penalised_model_selection} we show how the number of effective parameters should be used within model selection and sketch the derivation of a generalised model selection criterion for penalised-maximum-likelihood models. In Sec.~\ref{sec:results} we apply this generalised model selection criterion to our toy problem and show how it can be used to select the right order in the parametric approach or the optimal smoothing in the non-parametric models and how to choose between the two approaches. In Sec.~\ref{sec:bootstraps} we extensively discuss the efficiency of the method. The paper ends with a summary in Sec.~\ref{sec:summary}.

\section{A toy model based on Hermite Polynomials}
\label{sec:input}
\begin{figure}
    \centering
	\includegraphics[width=0.9\columnwidth]{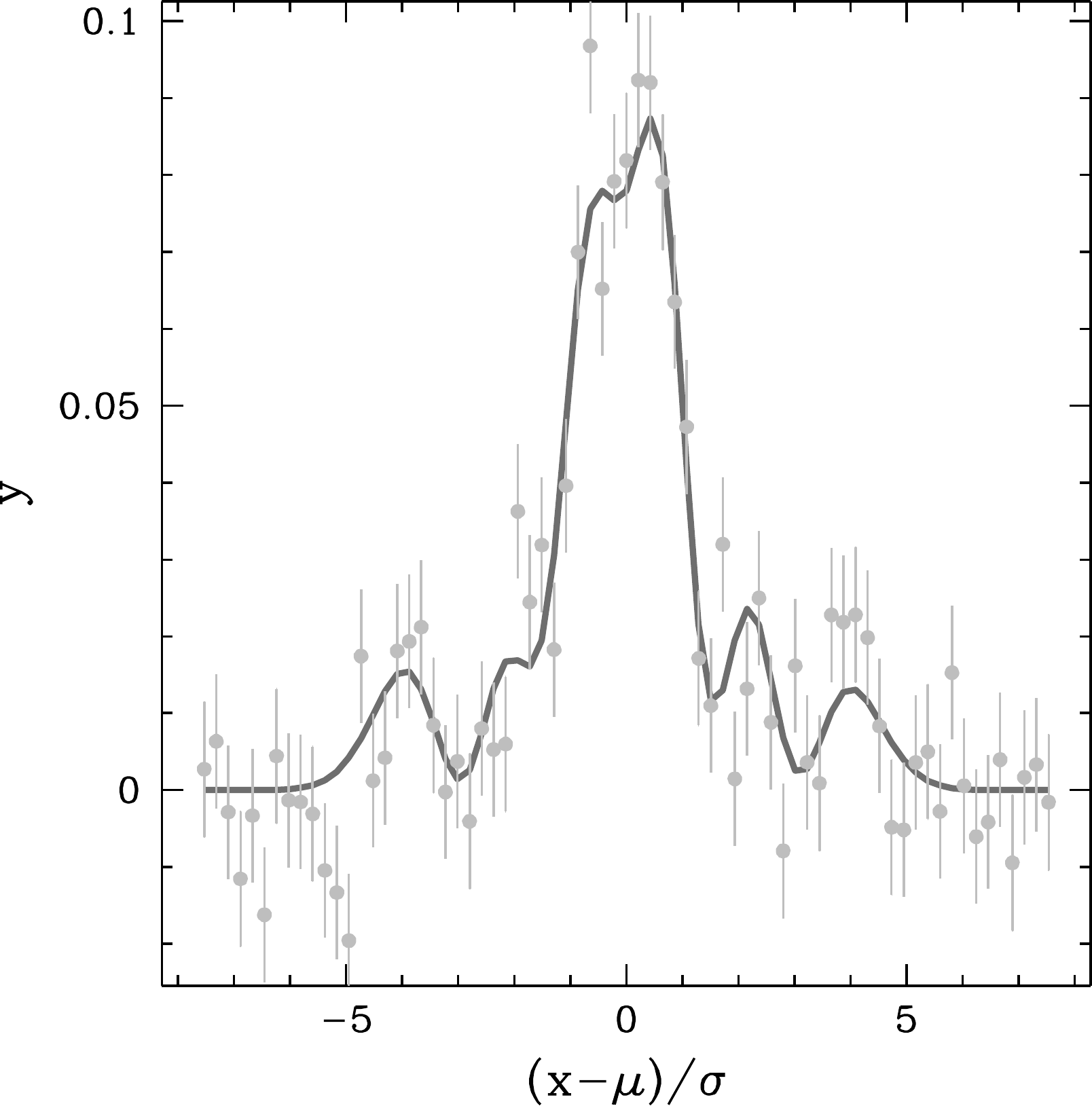}
    \caption{Example of simulated data $\dat_i = y(x_i)$ (eq.~\ref{eq:gh}). The solid grey line is the noise-free generating process $y_0(x)$. The gray dots simulate a noisy measurement, i.e. they represent a data sample $y$ obtained from $y_0$ by adding Gaussian noise (indicated by the error bars). The amount of noise is assumed to be constant along the x-axis. The SNR at the peak of the signal is 10. The generating input model is a Gauss-Hermite series up to order $\ngh = 10$ (cf. Tab.\ref{tab:input_table}). We create $\ndata=71$ data points evenly spaced between $\pm 8 \sigma$ of the Gauss component. }
    \label{fig:example_data}
\end{figure}

To illustrate the above outlined methods, we will try to recover a one dimensional function $y(x)$ from noisy mock observations. To this end we define $y_{0}(x)$ as a Gauss-Hermite series
\begin{equation}
\label{eq:gh}
y_0(x) = \frac{\gamma}{\sqrt{2 \pi \sigma}} \exp \left( - \frac{(x-\mu)^2}{2\sigma^2} \right) \, 
\left( 1+ \sum_{i=3}^{n_\mathrm{GH}} h_i \, H_i \left( \frac{x-\mu}{\sigma} \right) \right),
\end{equation}
where the $H_i$ are Hermite polynomials\footnote{Any finite and suitably smooth function $F(x)$ with $\lim x^3F(x) = 0$ for $x \to \pm \infty$ can be expanded into a Gauss-Hermite series \citep{myller07}. When ($\gamma,\mu,\sigma$) equal the parameters of the best fitting Gaussian function (we assume $\gamma$ and $\sigma$ are positive), then $h_1=1$ and $h_2=0$ and the series expansion can be written in the form of eq.~\ref{eq:gh} \citep[e.g.][]{vandermarel93}}. The noisy mock data $y(x)$ are obtained by adding Gaussian noise. The highest non-zero order of our generating model is $\ngh=10$ and the chosen coefficients are listed in Tab.~\ref{tab:input_table}. The resulting (noise-free) generating input model $y_0(x)$ is illustrated by the grey line in Fig.~\ref{fig:example_data}.

\begin{table}
	\centering
	\caption{The Gauss-Hermite coefficients of the generating (input) model. The underlying Gaussian function had $\mu=0$, $\sigma=350$ and $\gamma=1$. }
	\label{tab:input_table}
	\begin{tabular}{c|cccccccc} 
		\hline
		order $n$ & 3 & 4 & 5 & 6 & 7 & 8 & 9 & 10 \\
		\hline
        value & 0.0 &	0.1& 0.05 & 0.1 & -0.05 & 0.0 & 0.0 & 0.2 \\
	\end{tabular}
\end{table}

This toy model is inspired by the problem of fitting the shapes of gas emission lines in galaxies. In that case $y$ would represent a galaxy spectrum with the stellar continuum being subtracted and $x$ would be the logarithm of the wavelength. Usually, the shape of such emissions is close to a Gaussian but one could interpret the input model as a very complex emission line with several more or less separated Gaussian-like sub-components. However, for the purposes of this work one can also take this input model as an artificial mathematical model simply used to illustrate the capability of the model selection framework we propose.

\section{Two recovery methods: parametric vs non-parametric fits}
\label{sec:recovery}
We want to recover the generating model (Tab.~\ref{tab:input_table}) from the mock data in two different ways.

\subsection{Parametric Fits}
\label{subsec:parametric}
As our first set of models $\model(\modelparam)$ we use the Gauss-Hermite series of Eq.~\ref{eq:gh} itself, i.e. we fit $\model(\modelparam) = y(x;\modelparam)$. The parameters of this fit are $\modelparam=(\gamma,\mu,\sigma,h_3,h_4,\dots,h_n)$. We test different Gauss-Hermite models by varying $\ngh$, the maximum order included in the series. In the following only the case $\ngh \ll \ndata$ will be relevant and we call these fits parametric fits. 
We will derive the best-fit values of the $\ngh+1$ free parameters $\modelparam = (\gamma,\mu,\sigma,h_3,h_4,\dots,h_n)$ using a classical $\chi^2$ minimisation between the data $\dat$ and the model $\model(\modelparam)$,
\begin{equation}
\label{eq:chidef}
\chi^2 = \sum_{i=1}^{\ndata} \frac{(\dati-\model_i(\modelparam))^2}{\erri^2}
\end{equation}
where $\model_i(\modelparam) = \model(x_i;\modelparam)$.
Since the uneven Hermite polynomials are anti-symmetric with respect to the y-axis and the even ones are symmetric, we will always increase $\ngh$ in steps of two in our model fits.

\subsection{Non-parametric Fits}
\label{subsec:nonparametric}
In addition to the parametric fits, we will also perform non-parametric fits where our model $f$ consists of directly varying the $\model_i \equiv \modelparam_i$ of the signal at each of the $\ndata = 71$ values of $(x_i-\mu)/\sigma$ where a data point has been simulated. Again, we will determine the 71 free parameters $\modelparam_i$ of this model from a classical $\chi^2$ minimisation. The naive solution of this minimisation problem would be trivial, as a non-parametric model can fit all the data (including the noise) perfectly when $\model_i \equiv \modelparam_i$ = $\dat_i$, achieving $\chi^2 =0$. Therefore, in this non-parametric case, we also include a penalty term to penalise our non-parametric model against arbitrarily unsmooth solutions. As penalty we use the sum of the squared second derivatives of the non-parametric model with respect to $x$, i.e. we minimise $\chi^2 + \fsmooth \pfunc$ with
\begin{equation}
\label{eq:nonpara}
\pfunc =  \sum \left( \model_{i+1}-2\,\model_i+\model_{i-1} \right)^2.
\end{equation}
The factor $\fsmooth$ controls the strength of the smoothing constraints with $\fsmooth \to 0$ implying no smoothness enforced on the fits while $\fsmooth \to \infty$ implies strong smoothness constraints imposed on the model. By sampling different values of $\fsmooth$ we construct different non-parametric models from which we can select. 
For both, the parametric as well as the non-parametric fits, we use a standard Levenberg-Marquardt routine for the parameter estimation. 

\section{Model selection - Balancing Overfitting vs. Underfitting}
\label{sec:model_selection}

\begin{figure}
    \centering
	\includegraphics[width=0.9\columnwidth]{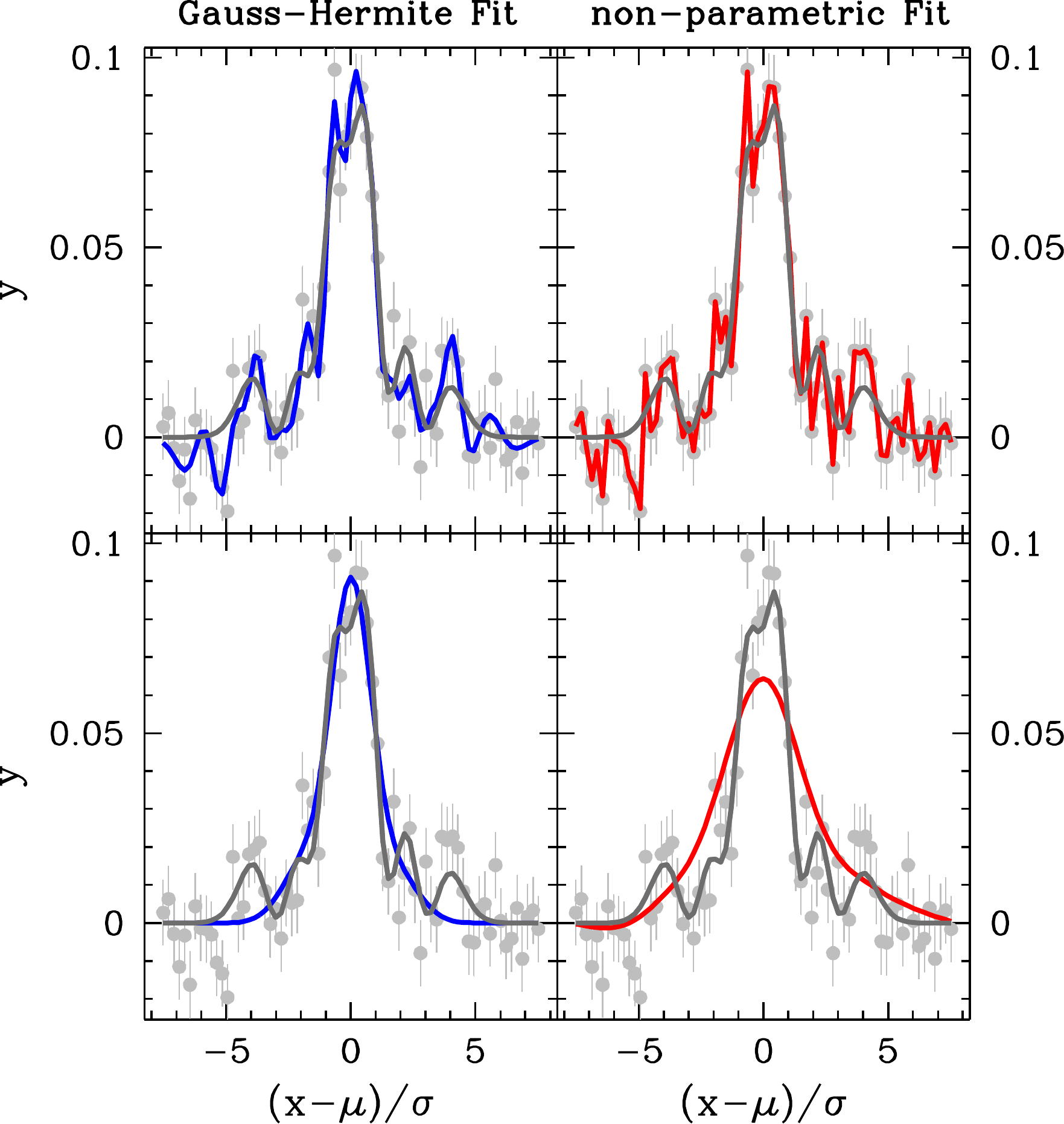}
    \caption{Example fits to the mock data of Fig.~\ref{fig:example_data}. The left panels show parametric fits with too many polynomial components ($\ngh=30$; top left panel) and with too few polynomial components ($\ngh=4$; bottom left panel). The right panels show non-parametric fits with a very low smoothing value ($\fsmooth=3 \, 10^8$; top right panel) and a very high smoothing value ($\fsmooth = 10^{11}$; bottom right panel). The generating input model is shown by the gray line in each panel together with the noisy mock data realisation (the same in each panel). The fits in the top panels do not recover the input model well because they {\it overfit} the data and follow the noise. The fits in the bottom panels do not recover the input model well either, but this time because they are {\it oversmooth} and not flexible enough to capture the structure of the generating model. }
    \label{fig:example_fits}
\end{figure}

Fig.~\ref{fig:example_fits} shows four example fits to the mock data set presented in Fig.~\ref{fig:example_data}. The fits in the left panels have been obtained with the parametric Ansatz. In the top-left panel $\ngh=30$ -- larger than in the generating model ($\ngh=10$, cf. Sec.~\ref{sec:input}). As a consequence, the model overfits the data and is too structured, compared to the generating input model. The opposite is true for the fit shown in the bottom-left panel, where $\ngh=4$. The resulting model is not flexible enough to capture all the structure of the generating model. 

The panels on the right hand side show the  analogous cases for the non-parametric fits. In the top-right panel the smoothing constraint is very weak such that the model fits the noise almost perfectly. In the bottom-right panel the model is instead oversmoothed and it can not fit the data appropriately. 

The goal of the following sections is to outline a simple method to find the optimum degree of model flexibility for both - the parametric and the non-parametric fit. The optimum degree of smoothness will be somewhere between the examples of overfitted and underfitted models shown in Fig.~\ref{fig:example_fits}. 

Model selection provides the general framework to chose between (not necessarily nested) models with different degrees of model flexibility. A commonly applied criterion used in this context is the Akaike information criterion (AIC, \citealt{Akaike74}). If $m$ is the number of free parameters in a fit and $\chi^2$ the achieved goodness-of-fit then a model is preferred when it has a lower 
\begin{equation}
    \label{eq:aic}
    \aic = \chi^2 + 2 \, m.
\end{equation}
The AIC originally comes from information theory and provides optimal model recovery in the sense that it aims to minimize the (estimated) Kullback-Leibler divergence -- a measure of the expected information loss -- between model and the actual generating process (cf. Sec.~\ref{sec:penalised_model_selection}). Intuitively, the AIC selects the model with the smallest number of variables $m$ among all models that provide a statistically viable fit, i.e. among all models with $\chi^2 + m \approx \ndata$. This is because for all these viable models, the AIC roughly reduces to $\aic=\ndata + m$ and, hence, becomes smallest for the lowest $m$. As such it's akin to Occam's razor.

Our intuitive understanding that the information loss in a fit tends to increase with the number of fitted parameters is often -- but not always -- sufficient to select an optimal model. An obvious requirement is that the number of parameters $m$ has to be known. This seems to be trivial for parametric fits where one usually knows every fit parameter explicitly and $m$ can be determined by simply counting the variables of the fitting function. However, even in the parametric case the determination of $m$ can become non-trivial when the model is strongly non-linear or when the parameters are not allowed to vary arbitrarily but instead are subject to constraint equations \citep{Andrae_2010}. 

The same applies to non-parametric models. Here, even more complications arise from smoothing penalties which often depend on one or several continuous smoothing parameters. Naively, the stronger the imposed smoothing constraints, the less flexible the model becomes. The invariable number $m$ of fitted parameters can therefore no longer be a measure of a model's flexibility or responsiveness to noise, respectively. This is important because noise -- by definition -- does not carry any information about the data generating structure. Hence it is actually a model's responsiveness to noise that determines the expected information loss. And if this is no longer encoded in $m$ then the simple model selection criterion of eq.~\ref{eq:aic} will no longer be sufficient. In fact, when penalties with a continuous parameter are present, the {\it effective} number of variables is expected to become a continuous variable as well. For example, in the above smoothing case: the stronger the smoothing the less flexible the model and the smaller the effective number of variables. In the next two sections we will demonstrate that the selection criterion eq.~\ref{eq:aic} can still be used under such circumstances, provided, however, that the generalised concept of the \textit{effective} number of parameters is applied to quantify a model's flexibility.

\section{Quantifying model flexibility - The number of effective free parameters}
\label{sec:meff}
In \citet{Lipka/Thomas2021} we first introduced such a generalized concept of \textit{effective} number of free parameters by using a flexible bootstrap method to estimate them. The following section is a brief review of the concepts first shown in \citet{Lipka/Thomas2021}.

Strictly speaking the number of free parameters $m$ is well defined only for parametric models that depend linearly on their free parameters \citep[e.g.][]{Hastie_2013} and have no a priori constraints imposed on their parameters \citep[e.g.][]{Andrae_2010}. However, the concept of free parameters can be generalized formally to more complex statistical models without relying on such restricting assumptions about the underlying model structure \citep[eg.][]{Ye_1998}. In such generalized frameworks the resulting degrees of freedom (i.e. the actual model flexibility) typically differs significantly from the number of variables of the fit model, and thus one can not derive the model flexibility by simply counting the number of variables. 

Therefore we employ bootstrap iterations to estimate the \textit{effective} number of free parameters (in the following $\meff$). To this end we establish a number of $\nboot$ bootstrap data sets $\bdat$ for each fit model by adding random Gaussian noise (based on the observed noise estimate $\err$) to an initial fit $\fit$ of said model to the observed data sample $\dat$. 
Thus, a set of bootstrap data $\bdat$ is generated at every data point $i$ by $\bootdat_i=\fit+{\cal N}(0,\erri)$ where ${\cal N}(0,\erri)$ is a Monte-Carlo realization drawn from the Gaussian distribution with mean $0$ and standard deviation $\err$. 
The goal of this bootstrap resampling technique is to emulate (or redraw) the observed data sample. 
As such bootstrapping assumes that the initial fit $\fit$ represents the (noise-free) data generating process well enough such that the bootstrap data can be treated as a resample of the observed sample\footnote{If this approximation is too crude, e.g. if the initial fit model is very different from the actual data generating process, then the bootstrapping estimation of $\meff$ is not legitimate. However, such models can easily be rejected anyways due to their overall bad fit to the data (e.g. if $\chi^2/\ndata\gg 1$).}. Each of the $\nboot$ sets of bootstrap data $\bdat^\iboot$ (for $\iboot = 1,...,\nboot$) are then fitted by the same model for which one attempts to estimate the flexibility, denoted as $\bfiti$.

The flexibility $\meff$ should be a measure of responsiveness of the model fit to noisy data. As such, a more flexible model should be able to follow more of the deviations in the data that was induced by the bootstrap noise than a less flexible model can. Therefore a 'natural' measure of this responsiveness is the normalized correlation of the model fit and the noisy data it was fitted to\footnote{Equation~\ref{eq:covariance_2} can be shown to be equivalent to the definition generalized degrees of freedom of \citet{Ye_1998}, which is a formal extension of the number of free parameters to non-linear models.}:
\begin{equation}
\meff=\sum_{i=1}^{\ndata}\frac{\cov\left(\model_i(\bootedparam),\bootdat_i \right)}{\erri^2} \leq \ndata
	\label{eq:covariance_2}
\end{equation}
Motivated by this expression (\ref{eq:covariance_2}) we introduce
\begin{equation}
\meffi=\sum_{i=1}^{\ndata} \left( \frac{\bfiti- \fiti}{\erri} \right) 
\left( \frac{\bdati- \fiti}{\erri} \right)
	\label{eq:covariance}
\end{equation}
where the sum goes over all data points $\ndata$ and $\iboot$ is the index of the bootstrap iteration. To reduce dependence on the specific noise realizations of the bootstrap data one should ideally average the results of eq.~\ref{eq:covariance} over multiple bootstrap iterations $\iboot=1,\ldots,\nboot$ such that
\begin{equation}
    \label{eq:meff}
    \meff = \ev(\meffi) \simeq \frac{1}{\nboot} \sum_{\iboot=1}^{\nboot} \meffi.
\end{equation}
Of course multiple iterations can be very computationally expensive if the fitting procedure in itself is complex. Fortunately, one can already achieve decent results without the need of that many bootstrap iterations as we will discuss in section~\ref{sec:bootstraps}.

\begin{figure}
    \centering
	\includegraphics[width=0.9\columnwidth]{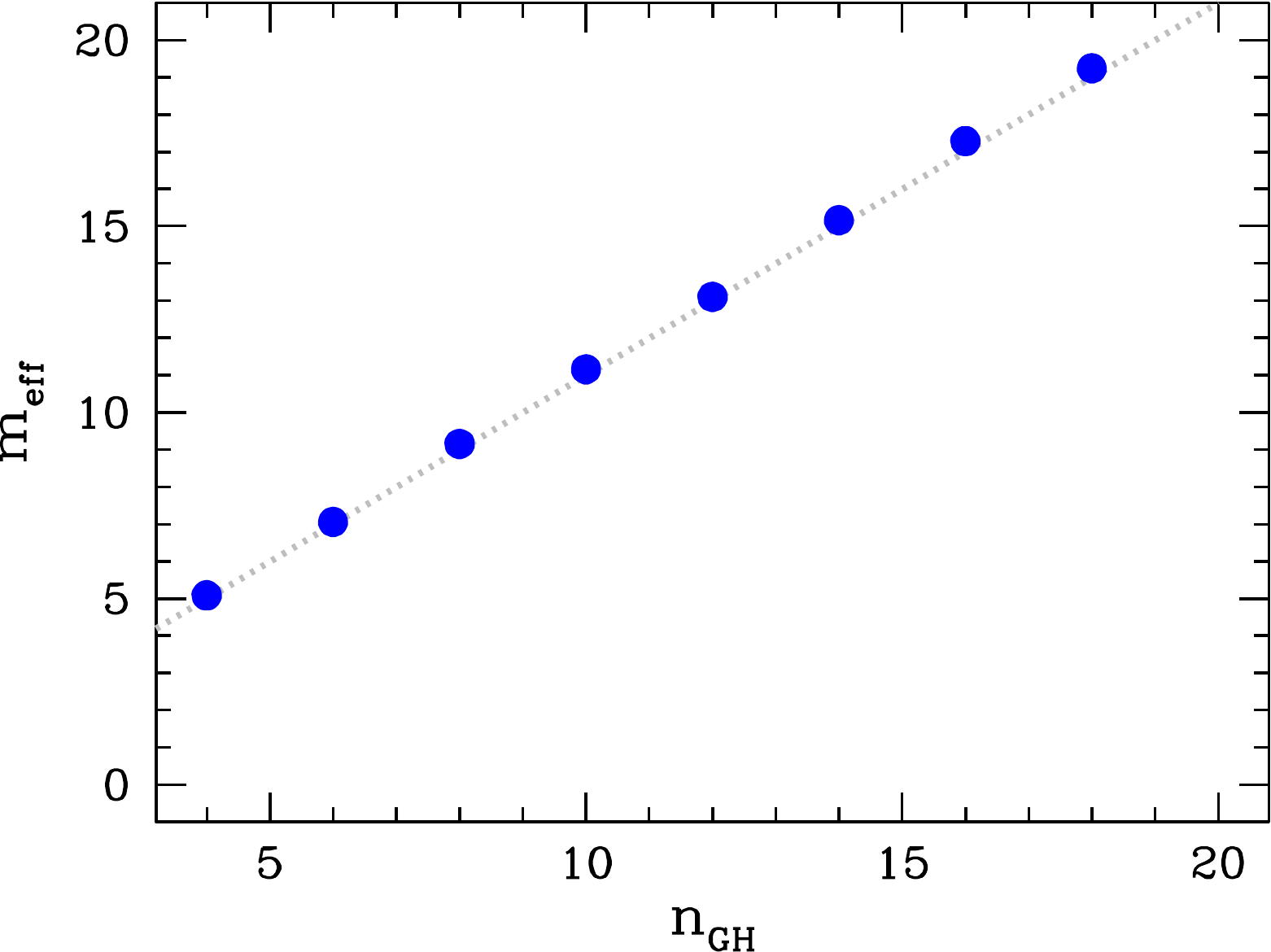}
    \caption{Estimated number of effective parameters $\meff$ for the parametric Gauss-Hermite models described in Sec.~\ref{subsec:parametric}. The blue dots show the numerically derived $\meff$ with $\nboot=500$. For the Gauss-Hermite models one can simply count the number of free parameters to $\meff = \ngh+1$ (dotted line). All the blue points fall exactly on the dotted line: the numerical evaluation of $\meff$ is very accurate.}
    \label{fig:example_meff}
\end{figure}

For linear models that are not subject to a penalty and for which the parameters are not restricted by equality/inequality constraints the number $m$ of fitted parameters happens to equal $\meff$ and therefore a simple count of the number of parameter $m$ is a viable measure for the responsiveness of the model to noise. We show this explicitly in App.~\ref{app:nopenalty}. Note that in standard linear theory $m$ actually only equals the naive number count of parameters if all parameters are linearly independent \citep[c.f.][]{Hastie_2013}. Nevertheless the equality of $\meff$ and $m$ (given aforementioned preconditions) still holds for dependent parameters as long as $m$ is calculated using standard linear theory instead of a naive counting of the number of parameters.

Our non-parametric fits provide an example for such a linear model (with $\partial f_i/\partial \modelparam_j = \delta_{ij}$) if no smoothing penalty is applied ($\fsmooth=0$). In that case the model will always yield $\chi^2=0$ with $f_j \equiv \modelparam_j=y_j$ such that 
\begin{equation}
\meff=\sum_{i=1}^{\ndata}\frac{\cov\left(\bootdat_i,\bootdat_i \right)}{\erri^2} = \ndata = m,
\end{equation}
as expected.

For non-linear models the equality $\meff=m$ might not hold in general, but under appropriate regularity conditions it will certainly hold locally. In practice, models which change their properties significantly over the uncertainty region of the data will in most cases not be very useful. Hence, if the model changes only slowly over the region in data space sampled by the bootstraps then $\meff = m$ will still hold (in the absence of penalties or parameter constraints). An example is the Gauss-Hermite series of section~\ref{subsec:parametric} where the parameter vector $\modelparam = (\gamma,\mu,\sigma,h_3,h_4,\dots,h_n)$ has a well-defined number of $\ngh+1$ elements, or free parameters respectively. Fig.~\ref{fig:example_meff} shows the numerically obtained $\meff$ of these non-linear models versus the (counted) number of free parameters. As expected, after $\nboot=500$ bootstraps the estimated effective number of parameters is equivalent to the counted number of free parameters.

In model selection the goal is to rank different models according to the estimated relative information loss between them. The classical $\aic=\chi^2+2m$ has been shown to be an unbiased estimator of this information loss, e.g. in the context of maximum-likelihood fits. However, we have motivated above that for penalised models, e.g., the ability of the model to adapt to noise depends on the strength of the penalty term and cannot be expressed by the invariable number $m$. By construction, $\meff$ is a more general measure of the model's flexibility, independent of the presence of a penalty or other constraints on the parameters. Therefore we argue that model selection in a more general context should involve $\meff$ rather than $m$ (the latter only being a measure of the responsiveness to noise under certain circumstances). The next Sec.~\ref{sec:penalised_model_selection} is dedicated to such a generalized model selection that extends to penalised models and, as such, will involve $\meff$ rather than $m$.

\section{Penalised likelihood: model selection with effective number of parameters}
\label{sec:penalised_model_selection}
In \citet{Lipka/Thomas2021}, within the context of orbit superposition models for galaxies, we have tested model selection techniques using $\meff$ rather than $m$. For a fixed gravitational potential the said orbit models are linear, but have a non-linear (maximum-entropy) penalty in our implementation \citep{Richstone88,Thomas04}. Comparing different weighting schemes $\chi^2 + w_m \meff$ we found that the AIC analogue (i.e. $w_m = 2$) performs best in estimating underlying properties of the data-generating processes. Therefore the results of \citet{Lipka/Thomas2021} suggest that the AIC can be generalised to penalised models by the substitution $m \rightarrow \meff$.

In the following we will motivate how model selection can be generalised for penalised models in a more formal way. For the reader who wants to skip this rather technical discussion
we preempt the important results of this Section: for penalised models, model selection indeed consists of minimising the generalised $\aicmod = \chi^2 + 2 \meff$. We derive this exactly for linear models. In Sec.~\ref{sec:results} we will apply the generalised $\aicmod$ to the toy model introduced in Sec.~\ref{sec:input}.

It is out of the scope of this paper to give a complete introduction to the foundation of model selection. A very good overview can be found in \citet{burnham02}. 
We simply start by recalling that Akaike model selection consists of minimising the \textit{expected}, \textit{estimated} information loss
\begin{equation}
    \label{eq:basiciloss}
             - \ev_\dat \ev_\bootdat (\logl (\bootdat|\fittedparam))
\end{equation}
(e.g. Chapter 7.2, \citealt{burnham02}). Here,
$\logl (\bootdat|\fittedparam))$ is the logarithm of the likelihood $\like$ of some fictitious data $z$ (see below) at the maximum-penalised likelihood estimate $\fittedparam$ of the model parameters $\modelparam$. The double expectation $\ev_y$ and $\ev_z$ deserve some further comments. The outer expectation $\ev_y$ is meant to reflect that -- conceptionally -- we aim at minimising the \textit{expected} information loss over large samples of actual data $y$. In the context of our toy model (Sec.~\ref{sec:input}) this would correspond to averaging the results over several mock data sets (we will come back to this in Sec.~\ref{subsec:model_correls}). However, in most practical applications one has only a single data set and needs an unbiased estimate of the information loss based on the actual data at hand (see below). The inner expectation $\ev_z$ reflects that the Kullback-Leibler divergence -- the measure of the information loss that underlies eq.~\ref{eq:basiciloss} -- is an integral that happens to have the \textit{form} of an expectation value. Hence, the integral over the integration variable $z$ can be expressed and interpreted as an expectation value over some fictitious data samples $z$. Below it will turn out that the bootstrap iterations we introduced in Sec.~\ref{sec:meff} are effectively the computation of the inner expectation $\ev_z$ over this fictitious data sample $z$ of eq.~\ref{eq:basiciloss}.

In the following -- for simplicity -- we restrict ourselves to a situation where the "truth" corresponds to one model among our candidates, i.e. there is a parameter vector $\modelparam_0$ of the "true" values of $\modelparam$\footnote{Model selection does not depend on the true model being among the candidates. While this complicates the discussion \citep[e.g.][]{burnham02} it does not change the conclusions in our context.}. If we would know $\modelparam_0$ we could use it in eq.~\ref{eq:basiciloss} to calculate the true information loss. However, in reality we only have an estimate $\fittedparam$ based on some noisy data $y$. Therefore eq.~\ref{eq:basiciloss} only quantifies the \textit{expected}, \textit{estimated} information loss based on $\fittedparam$. 

The standard derivation of AIC starts with the Taylor expansion
\begin{equation}
    \label{eq:taylor1}
    \begin{split}
    \logl(\bootdat|\fittedparam) \approx & \logl(\bootdat|\modelparam_0) + 
    \left[ 
    \frac{\partial \logl(\bootdat|\modelparam_0)}{\partial \Theta} 
    \right]^T 
    \left( \fittedparam-\modelparam_0\right) + \\
    & \frac{1}{2} \left( \fittedparam-\modelparam_0 \right)^T 
    \frac{\partial^2 \logl(\bootdat|\modelparam_0)}{\partial \Theta^2}
    \left( \fittedparam-\modelparam_0\right), \\
    \end{split}
\end{equation}
around this true parameter vector $\modelparam_0$. 
Here and in the rest of the paper we use the compact notation $\partial^2 \logl(\bootdat|\modelparam_0)/\partial \Theta^2$
to denote the Hessian matrix of $\logl$ evaluated at $\modelparam_0$.
For maximum-likelihood models the expectation of the linear term obviously vanishes. In our case of linear penalised models this holds true since $\ev_z(z) = f(\modelparam_0)$ (cf. App.~\ref{app:bootcondition}). Even if the truth is not among the candidate models this holds true under very weak conditions because $\ev_y(\fittedparam) = \modelparam_0$.

Then, using a second Taylor expansion the unknown $\logl(\bootdat|\modelparam_0)$ in eq.~\ref{eq:taylor1} is approximated as
\begin{equation}
    \label{eq:taylor2}
    \begin{split}
    \logl(\bootdat|\modelparam_0) \approx & \logl(\bootdat|\bootedparam) + 
    \left[ 
    \frac{\partial \logl(\bootdat|\bootedparam)}{\partial \modelparam} 
    \right]^T 
    \left( \modelparam_0 - \bootedparam\right) + \\
    & \frac{1}{2} \left( \modelparam_0-\bootedparam \right)^T 
    \frac{\partial^2 \logl(\bootdat|\bootedparam)}{\partial \modelparam^2}
    \left( \modelparam_0 -\bootedparam \right). \\
    \end{split}
\end{equation}
Within the classical maximum likelihood framework the linear term vanishes exactly for each $z$ -- by construction. The expectations of the remaining second order terms of eqs.~\ref{eq:taylor1} and \ref{eq:taylor2} can then be shown to combine to (the negative of) the number of fitted parameters $m$ which then leads to the classical form of the AIC \citep[e.g.][]{burnham02}.

In contrast, within a \textit{penalised} maximum likelihood framework, the linear term does not vanish and needs to be taken into account when combining eqs.~\ref{eq:taylor1} and \ref{eq:taylor2}. For linear models a simplification arises from the fact that $\partial^2 \logl/\partial \modelparam^2$ is a constant. Furthermore, since we will take the double expectation $\ev_y\ev_z$ and $\ev_y\ev_z(h(y))=\ev_y\ev_z(h(z))$ for any function $h$, we can substitute $\fittedparam$ by $\bootedparam$ in eq.~\ref{eq:taylor1}. Combining all the above, eq.~\ref{eq:basiciloss} becomes
\begin{equation}
    \label{eq:taylor5}
        \ev_y \ev_z(\logl(\bootdat|\fittedparam)) \approx  \ev_y \ev_z(\logl(\bootdat|\bootedparam)) +
    \ev_y \ev_z(h(z|\bootedparam)) 
\end{equation}
with
\begin{equation}
\begin{split}
    h(z|\bootedparam) & = \left[ 
    \frac{\partial \logl(\bootdat|\bootedparam)}{\partial \modelparam} 
    \right]^T 
    \left( \modelparam_0 - \bootedparam\right) + \\
    & \left( \modelparam_0-\bootedparam \right)^T 
    \frac{\partial^2 \logl(\bootdat|\bootedparam)}{\partial \modelparam^2}
    \left( \modelparam_0 -\bootedparam \right). \\
    \end{split}
\end{equation}
In our bootstrap simulations we use the estimate $\modelparam_0 \approx \fittedparam$. Then, using eqs.~\ref{eq:b19} and \ref{eq:b20} we find 
\begin{equation}
    \label{eq:motiv}
    \begin{split}
    \ev_z(h(z|\bootedparam)) = &
    \ev_z 
    \left[
    \left(\bootdat-f(\bootedparam)\right)^T \Sigma^{-1} \left(f(\fittedparam)-f(\bootedparam) \right) +  
    \right.
    \\
    & 
    \left.
    \left(f(\fittedparam)-f(\bootedparam) \right)^T \Sigma^{-1} \left(f(\fittedparam)-f(\bootedparam)\right)
    \right] 
    \end{split}
\end{equation}
and it is straight forward to show that this expectation value equals (the negative of) $\meff$ (cf. eq.~\ref{eq:mefflinear}). 

In the common situation where one cannot perform the expectation $\ev_y(\logl(\dat|\fittedparam))$ or, equivalently, $\ev_z(\logl(\bootdat|\bootedparam))$ one uses the estimate $\logl(\dat|\fittedparam)$, which for Gaussian errors reads $-\chi^2/2$. Taking everything together, model selection under penalised likelihood conditions then consists of minimising $\chi^2/2+\meff$
or, equivalently
\begin{equation}
    \label{eq:aicmod}
    \aicmod = \chi^2 + 2 \meff.
\end{equation}

It is natural to assume that the extended criterion of eq.~\ref{eq:aicmod} also holds (at least locally) for more general, non-linear models under appropriate regularity conditions (which to derive is out of the scope of this paper). In fact, in the next Sec.~\ref{sec:results} we will see that $\aicmod$ works equally well for our linear non-parametric model and for our non-linear Gauss-Hermite model. In the absence of a penalty (or of constraint equations for the parameters) $\meff = m$ (Sec.~\ref{sec:meff}) and $\aicmod=\aic$.

We note that the difference between $\aic$ and $\aicmod$ -- i.e. the replacement $m \to \meff$ -- arises because we do not assume that the models obey the maximum-likelihood condition in our derivation of $\aicmod$ as is the case for the classical $\aic$. Therefore $\aicmod$ encompasses penalised models as well. However, we want to stress that our derivation is independent of the specific conditions for \textit{penalised} maximum-likelihood models, meaning $\aicmod$ is not restricted to these models and may be applied in an even more general sense.

\begin{figure}
    \centering
	\includegraphics[width=0.9\columnwidth]{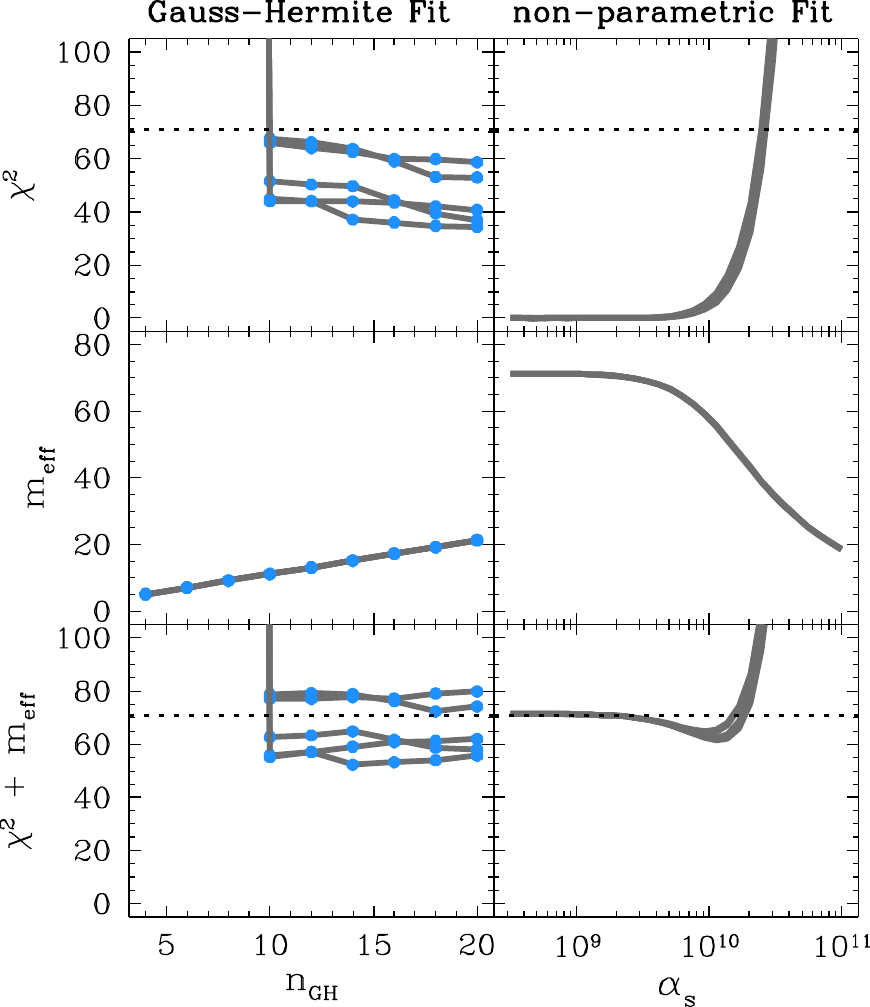}
	\caption{Illustration of varying model complexity in the Gauss-Hermite (parametric) fits (left panels) and the non-parametric fits (right panels). The generating model is the Gauss-Hermite series with $\ngh=10$ shown as the solid curve in Fig.~\ref{fig:example_data}. Each panel shows five gray lines, one for each of five different mock realisations of noisy data similar to the data shown in Fig.~\ref{fig:example_data} but assuming a SNR of 100 at the peak of the LOSVD. From top to bottom the plot shows the goodness of fit $\chi^2$, the number of effective parameters $\meff$ and $\chi^2 + \meff$. The number of data points, $\ndata=71$, is illustrated by the horizontal dotted lines. A statistically viable model must have a $\chi^2 + \meff \approx \ndata$. $\ngh$ is a discrete parameter of the parametric fits, its actually allowed values are highlighted in blue, the gray lines connecting the points have been added to better illustrate the trend. In the non-parametric case, $\fsmooth$ is a differentiable parameter and $\meff$ is a continuous function. For the parametric fits the model flexibility increases with the maximum order of the Gauss-Hermite fit, $\ngh$. For $\ngh\geq 10$ all fits become statistically viable. In the non-parametric case, the flexibility decreases as a function of the smoothing parameter $\fsmooth$, together with $\meff$. Below $\fsmooth \la 10^{10}$ all models lead to statistically acceptable fits. The middle-left panel confirms $\meff = \ngh+1$, i.e. the expected number of parameters for the parametric case.}
    \label{fig:chi_snr100}
\end{figure}

\section{Example results}
\label{sec:results}
Fig.~\ref{fig:chi_snr100} shows the result for both the parametric method (left panels) and the non-parametric method (right panels) in terms of fits to five different mock data realisations. In the parametric case, as expected, the number of fit variables ($\ngh+1$) increases with $\ngh$ and therefore the $\chi^2$ decreases with $\ngh$ (top-left and middle-left panels). In fact, as long as $\ngh<10$ -- i.e. when the number of fitted variables is smaller than in the generating model -- the fits do not yield a statistically viable fit to the data, because $\chi^2+\meff \gg \ndata$ (bottom-left panel). For $\ngh \ge 10$, while $\chi^2$ continues to decrease with increasing $\ngh$, all fits actually provide statistically equivalent representations of the data as $\chi^2 + \meff$ stays roughly constant in this regime. Implying the improvement in the goodness-of-fit $\chi^2$ is not significant but just as large as expected from the increased model flexibility.

\begin{figure}
    \centering
	\includegraphics[width=0.9\columnwidth]{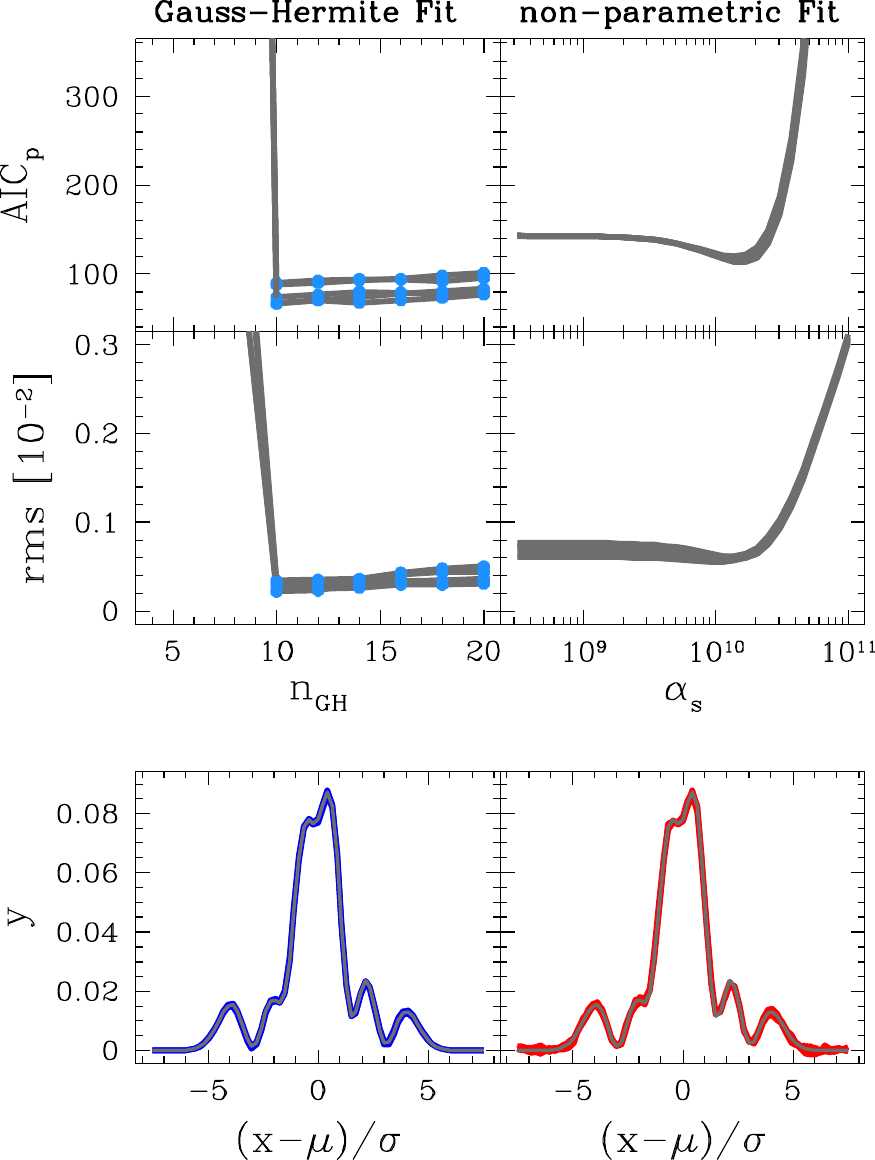}
    \caption{The same models as in Fig.~\ref{fig:chi_snr100}. The top panels show the $\aicmod = \chi^2 + 2 \, \meff$. The middle panels show the rms difference between the generating input model and the best-fit reconstruction of this input model from the fit to the mock data. Finally, the bottom panels show the generating model (gray) and the fitted reconstructions (blue/red) at the minimum $\aicmod$. The reconstruction of the input model is extremely good in both cases. In the parametric and in the non-parametric case the $\aicmod$ selection yields the model with the smallest rms difference to the generating input model.}
    \label{fig:aic_snr100}
\end{figure}

The non-parametric fits (right panels) behave similar, though $\meff$ increases from left to right (opposite to the parametric case) in this representation of $\fsmooth$. For low values of $\fsmooth$ the models are essentially unaffected by the smoothing penalty. As a result, each model variable $l_i$ becomes an entirely independent model variable and $\meff=\ndata$. In this regime, the model adapts perfectly to the noise in the data resulting in $\chi^2 \to 0$. The larger $\fsmooth$ the stronger the smoothing constraints become, meaning that $\meff$ continuously decreases whereas $\chi^2$ accordingly increases. Over a large interval of $\fsmooth$ this happens at constant $\chi^2+\meff \approx \ndata$. I.e. while the model becomes smoother it still leads to a statistically viable representation of the data. At some point, however, the smoothing constraints become so dominant that $\chi^2$ and $\chi^2+\meff$ increase significantly above $\ndata$. In that case the model becomes so dominated by the smoothing function that it can not yield a good representation of the data anymore. A noticeable difference between the parametric and the non-parametric fits is that the non-parametric $\chi^2$/$\chi^2+\meff$ curves appear to be much smoother than the parametric fits. We will come back to this in Sec.~\ref{subsec:model_correls}. 

Fig.~\ref{fig:aic_snr100} shows the $\aicmod$ (top panels), the recovery of the input model at the lowest $\aicmod$ (bottom panel) and the rms (root-mean-square)
between the generating input model and recovery from the fit (middle panels) -- again for both the parametric and the non-parametric fits.
For the rms we sum over the squared differences between the generating model and the fit at the $\ndata$ argument values of the data points.
Unsurprisingly, the $\aicmod$ of the parametric fits has a minimum at $\ngh=10$, the value used for the input model. For larger $\ngh$, even though the goodness-of-fit $\chi^2$ improves, the $\aicmod$ increases again because the models do not lead to a {\it significantly} better fit. This behaviour of the $\aicmod$ is mirrored exactly by the rms. The fact that the rms worsens with increasing $\ngh$ even though fits with $\ngh > 10$ are statistically viable is due to the fact that the models adapt more and more to the noise in the data, i.e. they start to overfit. The recovered LOSVDs at the optimum $\ngh$ agree very well with the input model.

\begin{figure}
	\centering
	\includegraphics[width=0.9\columnwidth]{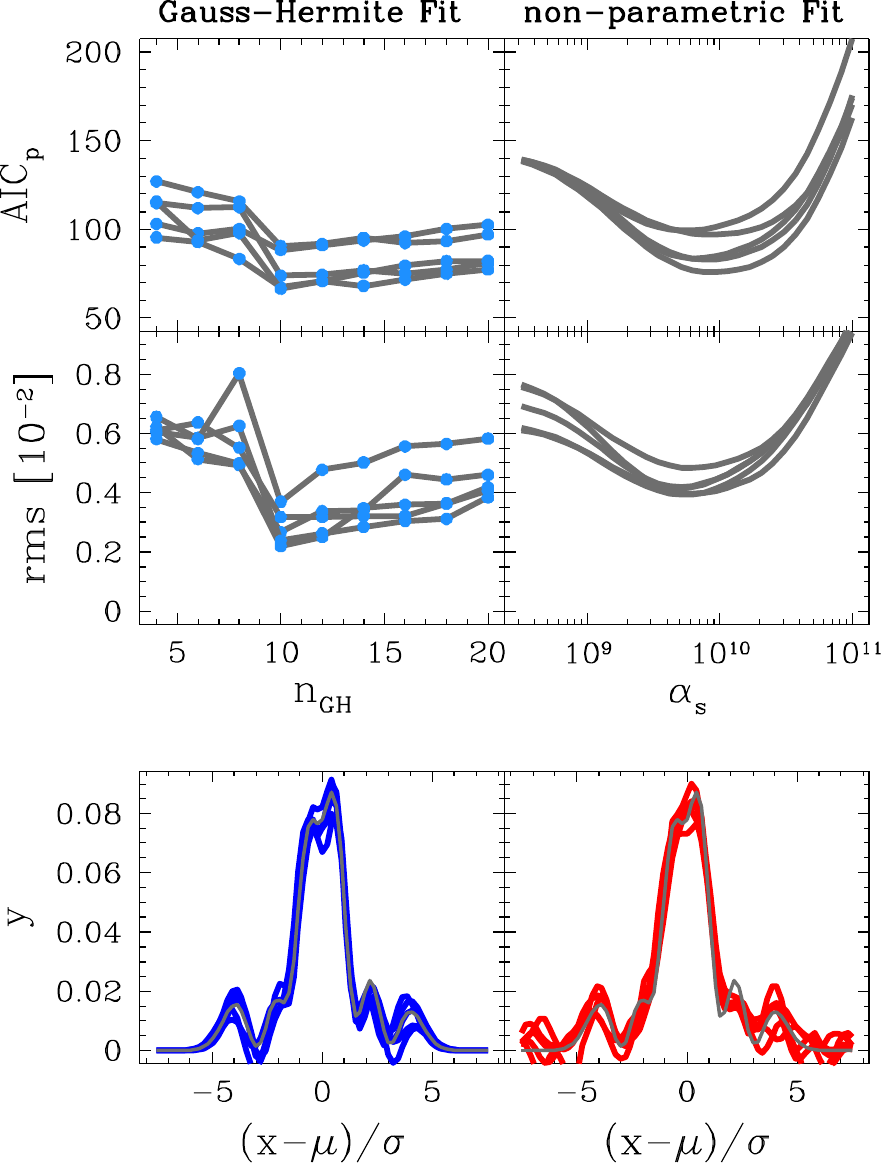}
    \caption{Same as Fig.~\ref{fig:aic_snr100} but the data are significantly noisier with an SNR=10. }
    \label{fig:aic_snr10_20ghmocks}
\end{figure}

The good recovery of the input model with the parametric fits is not that surprising since the data generating model (eq.~\ref{eq:gh}) is among the candidate models in this case. However, the recovery with the non-parametric models is almost equally good (right panels) even though in this case the generating model is not among the candidates. Again, the change of rms and the behaviour of the $\aicmod$ are very similar and the minimum $\aicmod$ is found to be where the recovery of the LOSVD is best. This shows that one can identify the optimal smoothing, or equivalently the optimum number of fit variables, even in the non-parametric case using the $\aicmod$ optimization and our definition of {\it effective} free parameters (Sec.~\ref{sec:meff}).

These above results and the fact that one can identify the model with the optimum number of fit variables from the $\aicmod$ and the $\meff$ does not depend on the assumed SNR. For the above fits, the SNR at the peak of the LOSVDs was set to SNR$=100$. Fig.~\ref{fig:aic_snr10_20ghmocks} shows the results for SNR$=10$. While the recovery of the LOSVDs gets more difficult due to the increased noise in the data, one can still identify the model with the optimum degree of model flexibility with the above described method.

\section{Evaluating the efficiency}
\label{sec:bootstraps}
It is common in non-parametric models to calibrate the smoothing by use of Monte Carlo simulations. With the above bootstrap approach such simulations are not necessary anymore. However, the efficiency of this approach will depend on the number of bootstrap iterations $\nboot$ necessary to obtain an accurate estimate of the optimum $\fsmooth$. 

In the $\aicmod$ framework the best choice for $\fsmooth$ follows from
\begin{equation}
    \label{eq:fsequation}
    \frac{\der \aicmod}{\der \fsmooth} = 0
\end{equation}
or
\begin{equation}
    \label{eq:dmequation}
    \frac{\der \chi^2}{\der \fsmooth} = -2 \frac{\der \meff}{\der \fsmooth},
\end{equation}
respectively. Therefore we actually only need an accurate estimate for $\der \meff/\der \fsmooth$, rather than for $\meff$ itself, to determine the optimum degree of smoothing $\fsmooth$. 

In this section we will compare the scatter in $\meff$ itself and the scatter of its derivative $\der \meff/\der \fsmooth$ with the goal to predict the number of bootstraps iterations $\nboot$ required to find the optimum $\fsmooth$.

\subsection{Bootstrap scatter in $\meff$}
\label{subsec:bootscat}

To estimate the scatter in $\meff$ it is convenient to define 
\begin{equation}
    \label{eq:X}
    a_i = \frac{\bfiti-\fiti}{\erri}
\end{equation}
and
\begin{equation}
    \label{eq:Y}
    b_i = \frac{\bdati-\fiti}{\erri}
\end{equation}
and treat them as random variables. The products $c_i = a_i b_i$ define the individual contributions to $\meffi$ (eq.~\ref{eq:covariance}) such that
\begin{equation}
    \label{eq:sumzA}
    \var (\meffi) = \var \left( \sum_i^{\ndata} c_i \right).
\end{equation}
From stochastic theory we can use Bienaymé's identity:
\begin{equation}
    \label{eq:sumzB}
    \var \left( \sum_i^{\ndata} c_i \right) = \sum_i^{\ndata} \var(c_i) + \sum_{i \ne j}^{\ndata} \cov(c_i,c_j)
\end{equation}
and
\begin{equation}
    \label{eq:varxy}
    \begin{split}
    \var(c_i) &= \cov(a_i^2,b_i^2) - 
    \left[ \cov(a_i,b_i) + \ev(a_i) \ev(b_i) \right]^2 + \\
    & \left( \var(a_i) + \ev(a_i)^2 \right)
    \left( \var(b_i) + \ev(b_i)^2 \right) \\
    \end{split}
\end{equation}
to evaluate eq.~\ref{eq:sumzA}.
By construction, $\ev(b_i) = 0$ and $\var(b_i) = 1$ such that eq.~(\ref{eq:varxy}) simplifies to
\begin{equation}
    \label{eq:varmeff}
    \var(c_i) = \cov(a_i^2,b_i^2) - 
    \cov(a_i,b_i)^2 + \var(a_i^2). 
\end{equation}
The quantities $a_i$ and $b_i$ (and $c_i$, respectively) can be calculated during the bootstrap iterations.
After all $\nboot$ bootstrap iterations, the scatter of $\meff$ is
\begin{equation}
    \label{eq:dmeff}
    \var(\meff) = \var(\meffi)/\nboot
\end{equation}
and can be estimated using the above equations.

The left panels of Fig.~\ref{fig:bootscatter} illustrate estimates of $\meff$ based on different $\nboot$ (solid lines) together with the calculated scatter $\errmeff = \sqrt{\var(\meff)}$ (dotted lines). We also include $\meff$ for $\nboot=2500$ as a reference (dashed line). Both, the estimate for $\meff$ and for its scatter improve with increasing $\nboot$ such that they can be determined with any desired accuracy. In practice, already after a relatively low number of bootstrap iterations $\nboot \sim 10$ the scatter from eq.~\ref{eq:dmeff} captures the uncertainty in $\meff$ very well and can be used to estimate the required $\nboot$. 

The exact behaviour of $\meff$ and $\errmeff$ will depend on the model function, the data and the smoothing function. However, in many cases -- as in Fig.~\ref{fig:bootscatter} -- the scatter $\errmeff$ can be presumed to increase with $\fsmooth$. Especially when the smoothing function biases the fit towards a single unique solution. If this favored reference model is not well chosen and far away from the true generating model (compared to $\erri$) the $\errmeff$ will be dominated by the $\ev(X_i^2)$ term (cf. eq.~\ref{eq:varxy}) leading to a large scatter in $\meff$. 

In our case, the smoothing function does not prefer a single unique set of values for the fitted $\fitpar$ because any straight line with any combination of slope/intercept will minimise the penalty function. Still, the scatter in $\meff$ increases noticeably with $\fsmooth$. As a rule of thumb we found that the scatter typically grows with $\chifs$, i.e. the goodness of fit of the original model $\fit$ at $\fsmooth$. Fortunately this also means that the regions with the largest scatter are typically not of interest anyways.
 
\subsection{Bootstrap scatter in $\dermeff$}
\label{subsec:bootscat_der}
As stated above, for the $\aicmod$ optimisation of $\fsmooth$ the scatter in $\der \meff/ \der \fsmooth$, and not in $\meff$, is the more important quantity. Therefore we will now evaluate the scatter in the derivative of $\meff$ with respect to $\fsmooth$. 

\begin{figure*}
	\centering
	\includegraphics[width=0.8\textwidth]{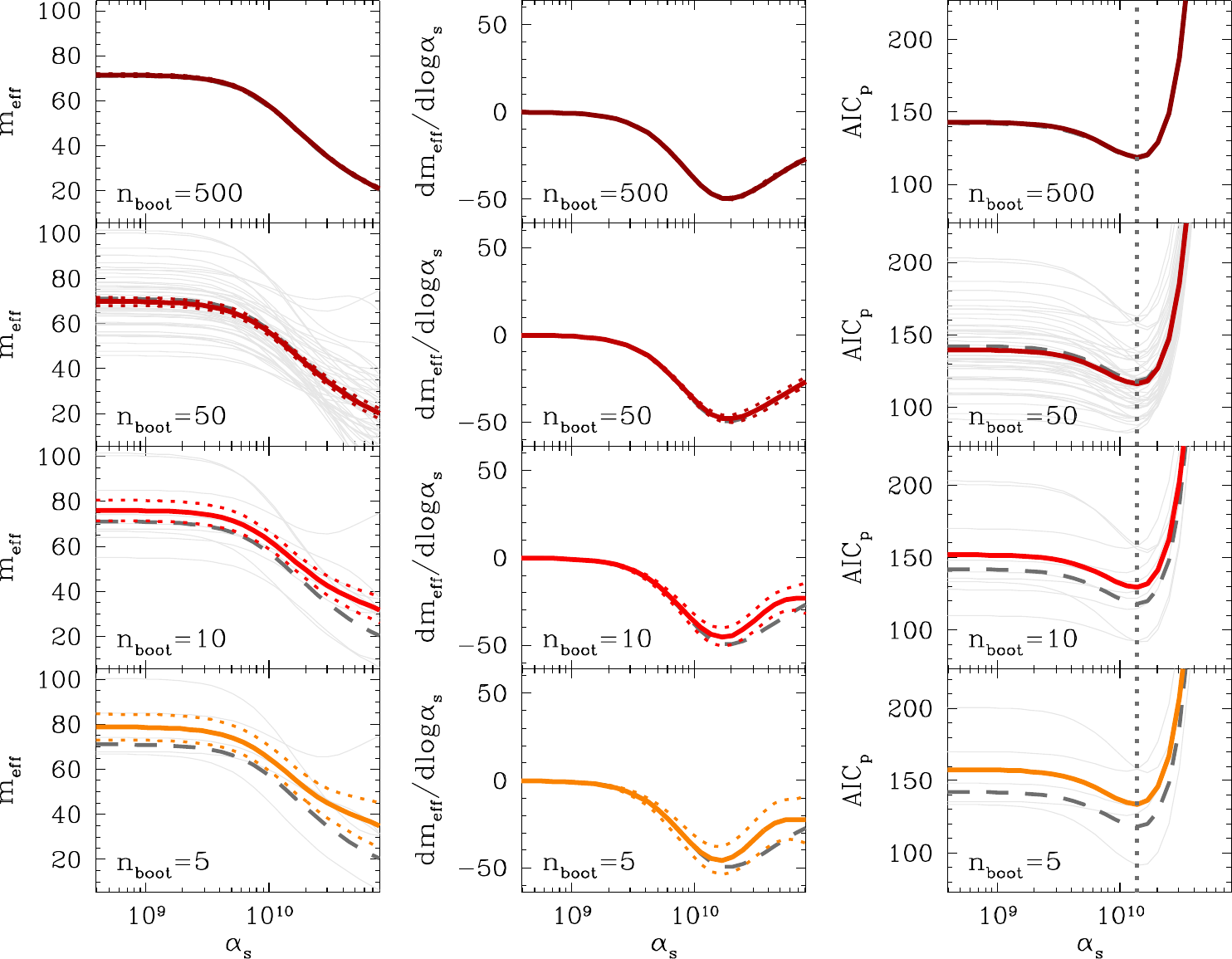}
    \caption{Scatter analysis of the models presented in Figs.~\ref{fig:chi_snr100} and \ref{fig:aic_snr100}. Left panels: number of effective parameters $\meff$ as a function of the smoothing factor $\fsmooth$ for different $\nboot$ (labelled in each panel). The colored solid curves show the mean $\meff$ over the respective $\nboot$, the dotted lines indicate $\errmeff$. For reference, the gray dashed line represents the case for $\nboot=2500$. In the three panels with $\nboot \le 50$ the individual $\meffi$ from each individual bootstrap iteration are included as well (thin solid lines). Most of the scatter in $\meff$ comes from individual $\meffi$ being scattered in the vertical direction as a whole in response to the particular noise pattern of each bootstrap iteration. The behaviour $\meffi(\fsmooth)$ depends very little on the noise pattern, in particular locally. Middle panels: similar as the left panels but the derivative $\logdermeff$ is shown instead of $\meff$. Rather than $\meff$ it is this derivative that is crucial to find the minimum of $\aicmod$. Because the scatter in $\meff$ mostly results from vertical shifts in the entire curves $\meffi(\fsmooth)$, the derivative $\logdermeff$ is very easy to compute with a higher accuracy, even with small $\nboot$. Right panels: $\aicmod$ as a function of $\fsmooth$ for different $\nboot$. The vertical dotted line indicates the value of $\fsmooth$ where the rms between in the generating input model and the reconstruction fit is smallest. For all the shown $\nboot$ it is correctly recovered by the minimum of $\aicmod$. In many situations even a single bootstrap iteration can be enough to get a decent optimisation of the smoothing in a non-parametric fit.}
    \label{fig:bootscatter}
\end{figure*}

Suppose we have two estimates of $\meff$ at two neighbouring values of $\fsmooth$, 
$\meff(\fsmooth)$ and $\meff(\fsmooth + \dfsmooth)$. The variance of the difference $\dmeffi = \meffi(\fsmooth + \dfsmooth)- \meffi(\fsmooth)$ is 
\begin{equation}
    \label{eq:vardmeffi}
    \begin{split}
    \var(\dmeffi) &= \var \left( \meffi(\fsmooth) \right) + \var \left( \meffi(\fsmooth + \dfsmooth) \right) - \\
    & 2 \, \cov \left( \meffi(\fsmooth), \meffi(\fsmooth + \dfsmooth) \right). \\
    \end{split}
\end{equation}
and, in analogy to $\meff$,
\begin{equation}
    \label{eq:vardmeff}
    \var(\dmeff) = \var(\dmeffi)/\nboot.
\end{equation}
What matters here -- beyond the scatter of $\meff$ itself -- is the correlation or covariance between neighbouring fits. For our penalty function (eq.~\ref{eq:nonpara}) the curves of the individual $\meffi(\fsmooth)$ are a smooth function of $\fsmooth$ (cf. Fig.~\ref{fig:bootscatter}). This is ensured if the penalty function is differentiable in $\fsmooth$, because then the correlation between neighbouring models is nearly maximal. E.g. if a specific noise pattern led to a bootstrap data set $\bdat$ that happened to result in a relative large $\meffi$ at $\fsmooth$ (compared to the mean $\meff(\fsmooth)$) then this will very likely also be true for $\meffi(\fsmooth + \dfsmooth)$. This holds locally if one uses the {\it same} noise pattern for neighbouring models. When comparing bootstrap fits at sufficiently different $\fsmooth$ this correlation will be weaker or might disappear completely (i.e. some of the $\meffi$ of Fig.~\ref{fig:bootscatter} cross).

In the middle column of panels of Fig.~\ref{fig:bootscatter} we plot $\dmeff(\fsmooth)$ for different $\nboot$ together with $\errdmeff = \sqrt{\var(\dmeff)}$. Due to the high degree of correlation between fits (and $\meff$) at neighbouring $\fsmooth$, the scatter in $\dmeff$ is very much reduced and even with less than $\nboot < 10$ bootstraps one can identify the characteristic behaviour of $\dmeff$.

Finally, the panels at the very right of Fig.~\ref{fig:bootscatter} show $\aicmod(\fsmooth)$ for different $\nboot$. For the optimisation of $\fsmooth$ the 
$\chifs$ term is significant as well. Similarly to $\meff$ we have
\begin{equation}
    \label{eq:chicovar}
    \begin{split}
    \var(\dchii) &= \var \left( \chii(\fsmooth) \right) + \var \left( \chii(\fsmooth + \dfsmooth) \right) - \\
    & 2 \, \cov \left( \chii(\fsmooth), \chii(\fsmooth + \dfsmooth) \right) \\
    \end{split}
\end{equation}
and basically all the above considerations about $\meff$ can be taken over to $\chifs$. As long as the penalty is differentiable in $\fsmooth$ the function $\chifs$ will be smooth (its actual behaviour is shown in the top-right panel of Fig.~\ref{fig:chi_snr100}). Consequently it is not surprising that the $\aicmod$ curves are very smooth even for very small $\nboot$. 

The vertical dotted line in the right panels of Fig.~\ref{fig:bootscatter} indicates the value of the smoothing factor $\fsmooth$ where the rms between the generating model and the fit has its minimum, i.e. the best model. In all the cases plotted in Fig.~\ref{fig:bootscatter} -- even for $\nboot$ as small as $\nboot=5$ -- this best model is correctly recovered by the $\aicmod$.

\subsection{Is a single bootstrap iteration enough to optimise $\fsmooth$?} 
\label{subsec:single_nbooy}
The small scatter in $\dermeff$ and $\chifs$ that results from the high degree of correlation between models with neighbouring $\fsmooth$ in case of a differentiable smoothing function make the model selection with $\meff$ a very efficient ansatz to optimise the smoothing in any kind of fit. 
In fact, in many cases already a single bootstrap iteration can be enough to get a decent estimate of $\fsmooth$.

\begin{figure}
	\centering
	\includegraphics[width=0.9\columnwidth]{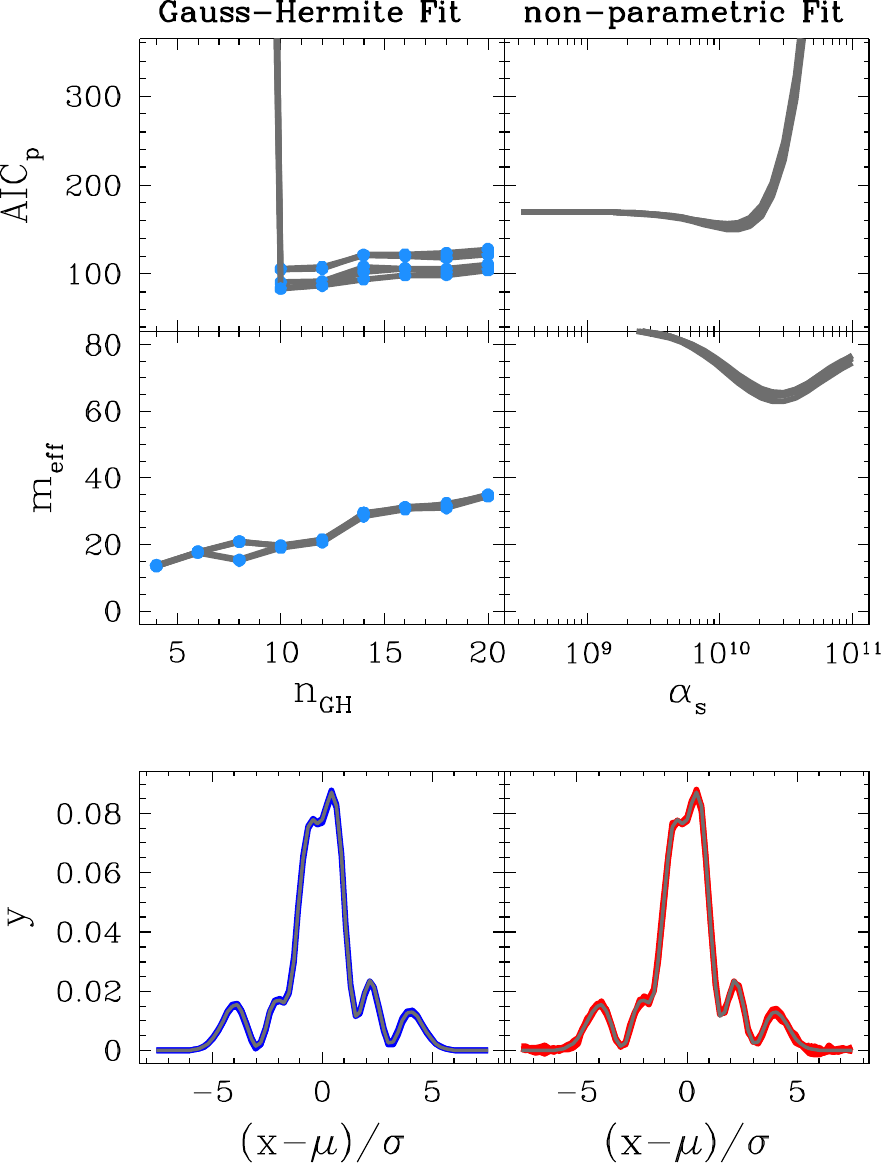}
    \caption{Similar to Figs.~\ref{fig:chi_snr100} and Fig.~\ref{fig:aic_snr100} but the number of bootstrap iterations to calculate $\meff$ has been reduced to $\nboot=1$. The Figure only shows $\aicmod$ and $\meff$. The recovery of the input model is almost not affected by the highly reduced number of bootstraps.}
    \label{fig:snr100_nboot1_combined}
\end{figure}

We illustrate this in Fig.~\ref{fig:snr100_nboot1_combined} which is similar to Figs.~\ref{fig:chi_snr100} and \ref{fig:aic_snr100} but the number of bootstrap iterations has been reduced to $\nboot=1$ and only $\aicmod$, $\meff$ and the reocvery of the input model are displayed. It is remarkable how well the recovery of the input model works:  after a single iteration, without any separate Monte Carlo simulations to calibrate $\fsmooth$, the recovery with the smallest rms is identified.

\subsection{Model Correlations}
\label{subsec:model_correls}
It is worth looking at the differences between the non-parametric case and the parametric case. The parametric analogue to the differentiable smoothing factor $\fsmooth$ is the order $\ngh$ of the parametric models. In contrast to $\fsmooth$, $\ngh$ is not a differentiable parameter. Rather, it is discrete. Hence, even fits with "adjancent" $\ngh$, i.e. fits at $\ngh$ and $\ngh+2$ are considerably different and less correlated than models with adjacent $\fsmooth$ are. This can be clearly seen from a comparison of the amount of scatter in the $\chi^2$ curves shown in the top panels of Fig.~\ref{fig:chi_snr100}. The discrete nature of $\ngh$ in the parametric models and the respective weaker correlation between models with similar but not identical $\ngh$ suppresses the covariance term in the analogue of eq.~\ref{eq:chicovar} for the parametric case. This leads to the jagged $\chi^2(\ngh)$ curves in the parametric case -- in contrast to the smooth $\chi^2(\fsmooth)$ in the non-parametric case. When the degree of correlation between the models is low, then {\it both} $\chi^2$ and $\meff$ become noisy.

As we have seen in Sec.~\ref{subsec:bootscat}, the noise in $\meff$ can be made arbitrarily small with a sufficiently large $\nboot$, i.e. by averaging over different noise patterns in the artificial {\it bootstrap} data. The noise in $\chi^2$ can be reduced in an analogous way, but this requires repeated measurements, i.e. averaging over different noise patterns in the {\it actual} data.

\begin{figure}
	\centering
	\includegraphics[width=1.0\columnwidth]{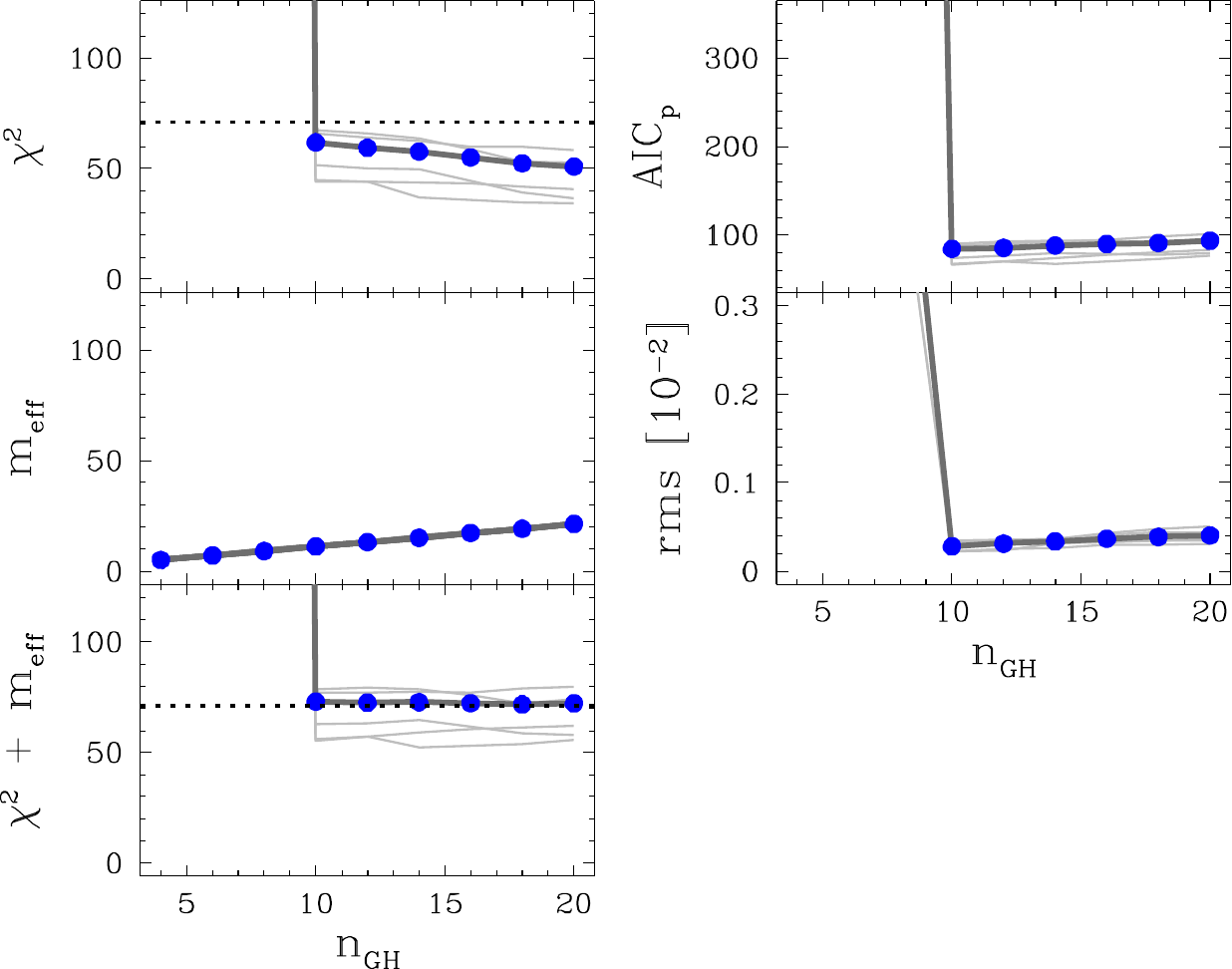}
    \caption{Left panels: same as the left panels in Fig.~\ref{fig:chi_snr100}, but the results have been averaged over 20 mock data sets. For comparison, the thin lines show again the results of the left panels in Fig.~\ref{fig:chi_snr100}. Right panels: same as the left panels in Fig.~\ref{fig:aic_snr100} but the results have been averaged over 20 mock data sets. As in the left panels, the results of Fig.~\ref{fig:aic_snr100} are shown for comparison as well (thin lines). This example shows how jagged $\chi^2$ curves that result from a lack of correlation between "neighbouring" models are smoothed out in the average over repeated measurements. (Such an average corresponds to the expectation $\ev_y$ in eq.~\ref{eq:basiciloss}.)}
    \label{fig:average_summary}
\end{figure}

This is illustrated in Fig.~\ref{fig:average_summary} where some of the parametric fit results that were already shown in Figs.~\ref{fig:chi_snr100} and \ref{fig:aic_snr100} are plotted again. However, in addition to the results for the five individual mock data sets, we also show the respective averages over fits to 20 mock data sets. As expected, these averages become perfectly smooth\footnote{In the context of eq.~\ref{eq:basiciloss} we here perform the expectation $\ev_y$.}.

Note that the non-parametric fits do not suffer from such a strong dependence on the mock noise since they depend in a differentiable way on the respective parameter $\fsmooth$.
The high degree of correlation between neighbouring models makes both the $\chi^2$ and the $\meff$ terms well behaved in this case. The model selection via $\meff$ is therefore very efficient to obtain the optimal smoothing in non-parametric models. 

\section{Summary}
\label{sec:summary}
We have introduced a simple data-driven method to optimise the smoothing of parametric and non-parametric models without the need of separate Monte-Carlo simulations. The method builds on a generalised concept of {\it effective} number of parameters \citep{Ye_1998,Lipka/Thomas2021} that can be easily computed for each model based on bootstrap simulations (cf. eq.~\ref{eq:meff}). It quantifies the complexity of a model in a very flexible way that can be used in linear as well as non-linear models, in models with or without constraint equations for the parameters and in models with or without penalties. In the simplest situation of a penalty-free model it reduces to the classical number of fitted parameters $m$. 

We have shown that the concept of effective number of parameters naturally emerges when the classical ideas of model selection are extended to models that do not fulfill the maximum-likelihood condition. In this case, the classical $\aic = \chi^2 + 2 m$ can be generalised to $\aicmod = \chi^2 + 2 \meff$. For linear models this holds exactly. The generalised $\aicmod$ can be applied to the large class of penalised maximum-likelihood models in particular.

As an application of the generalised model selection for penalised models we have tested two classes of fits to some mock data loosely inspired by the problem of fitting the shape of an emission line in a galaxy spectrum. The first model class is parametric and we showed how the generalised model selection leads to the recovery of the correct order of the fitting function. In this case the results of the generalised model selection and the classical AIC are supposed to coincide and they indeed do. In our second example, we showed how well the generalised model selection works to optimise the strength of a smoothing penalty in a non-parametric model. Note that for all the recoveries presented in this paper, the code was not provided a value for the smoothing factor but determined the optimal smoothing purely by itself and the data.

We have discussed in detail the efficiency of the method. While it does not require separate Monte-Carlo simulations to calibrate the optimal smoothing it requires bootstrap simulations to compute the effective number of parameters. A great advantage of the method is that for smooth model functions the number of required bootstraps is very low, of the order of 10 or even less. Each bootstrap represents a fit to a new -- bootstrapped -- data set, meaning that the extra-cost to optimise the smoothing is to do 10 rather than one fit per smoothing value.

The generalised model selection has probably many astrophysical applications.
In a previous paper we have already experimented with the generalised model selection in the context of orbit superposition models with tens of thousands of parameters and non-linear smoothing (entropy) constraints. There, the best generalised model selection criterion was the one derived here and model selection turned out necessary to obtain unbiased model results \citep{Lipka/Thomas2021}.

In a companion paper we will introduce a new spectral fitting code that makes use of the here discussed concepts to measure non-parametric line-of-sight velocity distributions of stars in galaxies (Thomas et al., in preparation). We anticipate that the method can be applied in many other situations like non-parametric deprojections \citep{Magorrian99,deNicola_2020} or any other situation where substructure has to be separated from noise like in strong gravitational lensing. Moreover, it is not tied to smoothing problems. The strength of any penalty function can be optimised in the same way as outlined here.

\section*{Acknowledgements}
We thank the anonymous referee for comments that helped improving the paper.

\section*{Data Availability}
The data underlying this article will be shared on reasonable request to the corresponding author.



\bibliographystyle{mnras}

\begin{thebibliography}{}
\makeatletter
\relax
\def\mn@urlcharsother{\let\do\@makeother \do\$\do\&\do\#\do\^\do\_\do\%\do\~}
\def\mn@doi{\begingroup\mn@urlcharsother \@ifnextchar [ {\mn@doi@}
  {\mn@doi@[]}}
\def\mn@doi@[#1]#2{\def\@tempa{#1}\ifx\@tempa\@empty \href
  {http://dx.doi.org/#2} {doi:#2}\else \href {http://dx.doi.org/#2} {#1}\fi
  \endgroup}
\def\mn@eprint#1#2{\mn@eprint@#1:#2::\@nil}
\def\mn@eprint@arXiv#1{\href {http://arxiv.org/abs/#1} {{\tt arXiv:#1}}}
\def\mn@eprint@dblp#1{\href {http://dblp.uni-trier.de/rec/bibtex/#1.xml}
  {dblp:#1}}
\def\mn@eprint@#1:#2:#3:#4\@nil{\def\@tempa {#1}\def\@tempb {#2}\def\@tempc
  {#3}\ifx \@tempc \@empty \let \@tempc \@tempb \let \@tempb \@tempa \fi \ifx
  \@tempb \@empty \def\@tempb {arXiv}\fi \@ifundefined
  {mn@eprint@\@tempb}{\@tempb:\@tempc}{\expandafter \expandafter \csname
  mn@eprint@\@tempb\endcsname \expandafter{\@tempc}}}

\bibitem[\protect\citeauthoryear{Akaike}{Akaike}{1973}]{Akaike73}
Akaike H.,  1973, Information Theory and an Extension of the Maximum Likelihood
  Principle.
Springer New York, New York, NY, pp 199--213

\bibitem[\protect\citeauthoryear{{Akaike}}{{Akaike}}{1974}]{Akaike74}
{Akaike} H.,  1974, IEEE Transactions on Automatic Control, \href
  {https://ui.adsabs.harvard.edu/abs/1974ITAC...19..716A} {19, 716}

\bibitem[\protect\citeauthoryear{{Andrae}, {Schulze-Hartung}  \&
  {Melchior}}{{Andrae} et~al.}{2010}]{Andrae_2010}
{Andrae} R.,  {Schulze-Hartung} T.,   {Melchior} P.,  2010, arXiv e-prints,
  \href {https://ui.adsabs.harvard.edu/abs/2010arXiv1012.3754A} {p.
  arXiv:1012.3754}

\bibitem[\protect\citeauthoryear{Burnham \& Anderson}{Burnham \&
  Anderson}{2002}]{burnham02}
Burnham K.,  Anderson D.,  2002, Model selection and multimodel inference: a
  practical information-theoretic approach.
Springer Verlag

\bibitem[\protect\citeauthoryear{{\noopsort{De Nicola}}{de Nicola}, {Saglia},
  {Thomas}, {Dehnen}  \& {Bender}}{{\noopsort{De Nicola}}{de Nicola}
  et~al.}{2020}]{deNicola_2020}
{\noopsort{De Nicola}}{de Nicola} S.,  {Saglia} R.~P.,  {Thomas} J.,  {Dehnen}
  W.,   {Bender} R.,  2020, \mn@doi [\mnras] {10.1093/mnras/staa1703}, \href
  {https://ui.adsabs.harvard.edu/abs/2020MNRAS.496.3076D} {496, 3076}

\bibitem[\protect\citeauthoryear{Hastie, Tibshirani  \& Friedman}{Hastie
  et~al.}{2013}]{Hastie_2013}
Hastie T.,  Tibshirani R.,   Friedman J.,  2013, The Elements of Statistical
  Learning - Data Mining, Inference, and Prediction.
Springer Science \& Business Media, Berlin Heidelberg

\bibitem[\protect\citeauthoryear{{Lipka} \& {Thomas}}{{Lipka} \&
  {Thomas}}{2021}]{Lipka/Thomas2021}
{Lipka} M.,  {Thomas} J.,  2021, \mn@doi [\mnras] {10.1093/mnras/stab1092},
  \href {https://ui.adsabs.harvard.edu/abs/2021MNRAS.504.4599L} {504, 4599}

\bibitem[\protect\citeauthoryear{{Magorrian}}{{Magorrian}}{1999}]{Magorrian99}
{Magorrian} J.,  1999, \mn@doi [\mnras] {10.1046/j.1365-8711.1999.02135.x},
  \href {https://ui.adsabs.harvard.edu/abs/1999MNRAS.302..530M} {302, 530}

\bibitem[\protect\citeauthoryear{{Myller-Lebedeff}}{{Myller-Lebedeff}}{1907}]{myller07}
{Myller-Lebedeff} W.,  1907, Mathematische Annalen, 64, 388

\bibitem[\protect\citeauthoryear{{Richstone} \& {Tremaine}}{{Richstone} \&
  {Tremaine}}{1988}]{Richstone88}
{Richstone} D.~O.,  {Tremaine} S.,  1988, \mn@doi [\apj] {10.1086/166171},
  \href {https://ui.adsabs.harvard.edu/abs/1988ApJ...327...82R} {327, 82}

\bibitem[\protect\citeauthoryear{{Thomas}, {Saglia}, {Bender}, {Thomas},
  {Gebhardt}, {Magorrian}  \& {Richstone}}{{Thomas} et~al.}{2004}]{Thomas04}
{Thomas} J.,  {Saglia} R.~P.,  {Bender} R.,  {Thomas} D.,  {Gebhardt} K.,
  {Magorrian} J.,   {Richstone} D.,  2004, \mn@doi [\mnras]
  {10.1111/j.1365-2966.2004.08072.x}, \href
  {https://ui.adsabs.harvard.edu/abs/2004MNRAS.353..391T} {353, 391}

\bibitem[\protect\citeauthoryear{{\noopsort{Van der Marel}}{van der Marel} \&
  {Franx}}{{\noopsort{Van der Marel}}{van der Marel} \&
  {Franx}}{1993}]{vandermarel93}
{\noopsort{Van der Marel}}{van der Marel} R.~P.,  {Franx} M.,  1993, \mn@doi
  [\apj] {10.1086/172534}, \href
  {https://ui.adsabs.harvard.edu/abs/1993ApJ...407..525V} {407, 525}

\bibitem[\protect\citeauthoryear{Ye}{Ye}{1998}]{Ye_1998}
Ye J.,  1998, \mn@doi [Journal of the American Statistical Association]
  {10.1080/01621459.1998.10474094}, 93, 120

\makeatother
\end{thebibliography}

\providecommand{\noopsort}[1]{}




\appendix

\section{The equivalence of $\mcount$ and $\meff$ in models without a penalty}
\label{app:nopenalty}
The bootstrap data are constructed by adding noise to the best-fit model obtained from the actual data. The $\chi^2$ of the bootstrapped data with respect to the original model reads
\begin{equation}
    \chiprior = \sum_{i=1}^{\ndata} \left( \frac{\bdati-\fiti}{\erri} \right)^2. 
\end{equation}
We call this quantity $\chiprior$ because it represents the residuals prior to the bootstrap fit. $\chiprior$ follows a $\chi^2$ distribution with $\ndata$ degrees of freedom, i.e. $\ev(\chiprior)=\ndata$. 

After the bootstrap fit, the $\chi^2$ with respect to the best-fit model obtained from the bootstrapped data reads
\begin{equation}
    \chipost = \sum_{i=1}^{\ndata} \left( \frac{\bdati-\bfiti}{\erri} \right)^2.
\end{equation}
To contrast it from the above $\chiprior$ we call it $\chipost$ because it represents the residuals posterior to the bootstrap fit.
For a linear model with $m$ independent variables $\chipost$ follows a $\chi^2$ distribution with $\ndata-m$ degrees of freedom implying $\ev(\chipost)=\ndata-m$.


With some simple algebraic conversions, Eq.~\ref{eq:meff} can be written as
\begin{equation}
    \label{eq:dchi}
    \begin{split}
    \ev(\meffi) = &\ev(\chiprior)-\ev(\chipost) + \\
    & \sum_{i=1}^{\ndata} \ev \left[ \left( \frac{\bfiti-\bdati}{\erri} \right)
    \left( \frac{\bfiti-\fiti}{\erri} \right) \right]
    \end{split}
\end{equation}
and further transformed into
\begin{equation}
    \label{eq:dchi2}
    \begin{split}
    \meff = & \ev(\chiprior)-\ev(\chipost)   + \\ 
    & \sum_{i=1}^{\ndata} \ev \left[ \left( \frac{\bfiti-\bdati}{\erri^2} \right) \bfiti \right] - \\
    & \sum_{i=1}^{\ndata} \ev \left[ \left( \frac{\bfiti-\bdati}{\erri^2} \right) \fiti \right].
    \end{split}
\end{equation}
In the absence of a penalty $\pfunc$ the mean of the fit is invariant under the bootstrap iterations, $\ev(\bfiti) = \fiti$ (App.~\ref{app:bootcondition}). Since by construction $\ev(\bdati) = \fiti$ the last sum of expectation values therefore vanishes. Moreover, without a penalty function it can also be shown that 
\begin{equation}
    \sum_{i=1}^{\ndata} \ev \left( \frac{\bdati\bfiti }{\erri^2} \right) 
    =
    \sum_{i=1}^{\ndata} \ev \left( \frac{\bfiti \bfiti }{\erri^2} \right)
\end{equation}
(App.~\ref{app:booteq}) meaning that also the first sum of expectation values in eq.~\ref{eq:dchi2} vanishes.

Hence, for linear models without constraints and penalties 
\begin{equation}
    \label{eq:meffm}
    \meff = \ev(\chiprior)-\ev(\chipost) = m.
\end{equation}

This means that in the absence of a penalty term, $\meff$ behaves exactly as the classical number of variables $m$. In this case, $m$ \textit{is} a measure of the responsiveness of the model to noise. The equality is no longer guaranteed when the parameter estimation is subject to a penalty term.

\section{Specific properties of linear models}
\label{app:bootcondition}
A linear model with $m$ parameters can be represented by a matrix $\matr{A}$ and the parameter vector $\modelparam=(\param_1,\ldots,\param_m)$ such that the model vector $f=\matr{A} \modelparam$ or, in index notation,
\begin{equation}
    \model_i = \sum_{k=1}^{m} A_{ik} \modelparam_k.
\end{equation}
The matrix $\matr{A}$ consists of the partial derivatives of the model $\model$ with respect to the parameters $\modelparam$,
\begin{equation}
    \frac{\partial \model_i}{\partial \modelparam_k} = A_{ik}.
\end{equation}

For simplicity we assume uncorrelated Gaussian errors such that the log-likelihood of the model reads $\logl \sim -\chi^2/2$ where
\begin{equation}
    \chi^2 = (y-\matr{A}\modelparam)^T \matr{\Sigma}^{-1} (y-\matr{A}\modelparam).
\end{equation}
Here, $y$ is the data vector
and $\matr{\Sigma}$ is the variance-covariance matrix.

At the fitted parameter values $\fittedparam$ the model reads
\begin{equation}
    \fiti = \sum_{k=1}^{m} A_{ik} \fittedparamk
\end{equation}
and the bootstrap can be written as
\begin{equation}
    \bdati = \sum_{k=1}^{m} A_{ik} \fittedparamk + \bnoisei
\end{equation}
where $\bnoisei$ is the bootstrap noise of iteration $\iboot$. Finally, 
\begin{equation}
    \bfiti = \sum_k A_{ik} \bootedparamkiota.
\end{equation}

The $\chi^2$ minimisation implies
\begin{equation}
    \label{eq:minimise}
     \sum_{i=1}^{\ndata} \left( \frac{2 (\bdati-\bfiti)}{\erri^2} \frac{\partial \model_i}{\partial \param_j} \left( \bootedparamiota \right) \right) = 0.
\end{equation}

which for the linear models translates into the $m$ equations
\begin{equation}
    \sum_{i=1}^{\ndata} \frac{2}{\erri^2} \left( 
    \sum_{k=1}^{m} A_{ik}(\fittedparamk-\bootedparamkiota) + \bnoisei
    \right) A_{ij} = 0
\end{equation}
for $j=1,\ldots,m$.
Taking the expectation value of the above over many bootstrapped data samples $z$ we can simplify using $\ev(\bnoisei)=0$ and with
\begin{equation}
    B_{kj} = \sum_{i=1}^{\ndata} {\cal A}_{ki}^T {\cal A}_{ij}
\end{equation}
and ${\cal A}_{ij} = A_{ij}/\epsilon_i$ these m equations read
\begin{equation}
    \sum_{k=1}^{m} B_{kj} \ev(\fittedparamk - \bootedparamkiota) = 0.
\end{equation}
When the $m$ variables are independent the matrix $\matr{A}$ has maximum rank and $\matr{B}$ is a $m \times m$ matrix of rank $m$. Then the bootstrap assumption $\ev(\bfiti) = \fiti$ follows because the expectation values $\ev(\fittedparamk - \bootedparamkiota)$ have to be zero and $\ev(\bootedparamkiota) = \fittedparamk$ implies  $\ev(\bfiti) = \fiti$.

\label{app:booteq}
Multiplying the $m$ equations of the maximum-likelhood condition 
\begin{equation}
    \sum_{i=1}^{\ndata} \left( \frac{2 (\bdati-\bfiti)}{\erri^2} \frac{\partial \model_i}{\partial \param_j} \left( \bootedparamiota \right) \right) = 0,
\end{equation}
$j=1,\ldots,m$, each by $\bootedparamjiota$
\begin{equation}
    \sum_{i=1}^{\ndata} \left( \frac{ (\bdati-\bfiti)}{\erri^2} \frac{\partial \model_i}{\partial \param_j} \left( \bootedparamiota \right) \right) \bootedparamjiota = 0
\end{equation}
then
\begin{equation}
    \sum_{i=1}^{\ndata} \sum_{j=1}^{m} \left( \frac{ (\bdati-\bfiti)}{\erri^2} \frac{\partial \model_i}{\partial \param_j} \left( \bootedparamiota \right) \right) \bootedparamjiota = 0.
\end{equation}
and, thus, for linear models
\begin{equation}
    \sum_{i=1}^{\ndata} \left( \frac{ \bdati-\bfiti}{\erri^2} \right) \bfiti  = 0
\end{equation}
which means that
\begin{equation}
    \label{eq:noname}
    \sum_{i=1}^{\ndata} \ev \left( \frac{\bdati\bfiti }{\erri^2} \right) 
    =
    \sum_{i=1}^{\ndata} \ev \left( \frac{\bfiti \bfiti }{\erri^2} \right).
\end{equation}

Under the condition of a penalised maximum likelihood the
penalty function ${\cal P}$ modifies eq.~\ref{eq:minimise} to
\begin{equation}
    \sum_{i=1}^{\ndata} \left( \frac{2 (\bdati-\bfiti)}{\erri^2} \frac{\partial \model_i}{\partial \param_j} \left( \bootedparamiota \right) \right)  -  \fsmooth \frac{\partial \pfunc}{\partial \param_j} =0.
\end{equation}
Consequently, even for linear models neither $\ev(\bfiti) = \fiti$ nor eq.~\ref{eq:noname} can be assumed to hold in this case.

For linear models as above 
\begin{equation}
    \frac{\partial \logl(y|\modelparam)}{\partial \modelparam} = (y-\matr{A}\modelparam)^T \matr{\Sigma}^{-1} \matr{A}
\end{equation}
and 
\begin{equation}
    \frac{\partial^2 \logl(y|\modelparam)}{\partial \modelparam^2} = - \matr{A}^T \matr{\Sigma}^{-1} \matr{A}.
\end{equation}
For some parameter vector $\tilde{\modelparam}$ this implies
\begin{equation}
    \label{eq:b19}
    \left( \frac{\partial \logl(y|\modelparam)}{\partial \modelparam} \right) \tilde{\modelparam} = 
    (y-f(\modelparam))^T \matr{\Sigma}^{-1} f(\tilde{\modelparam})
\end{equation}
and
\begin{equation}
    \label{eq:b20}
    \tilde{\modelparam}^T \left( \frac{\partial^2 \logl(y|\modelparam)}{\partial \modelparam^2} \right) \tilde{\modelparam}
    = - f(\tilde{\modelparam})^T \matr{\Sigma}^{-1} f(\tilde{\modelparam}). 
\end{equation}
In particular, the last expression does not depend on $\modelparam$ but only on $\tilde{\modelparam}$.

For linear models, the definition of $\meff$ (cf. eq.~\ref{eq:covariance}) can be written as
\begin{equation}
    \label{eq:mefflinear}
    \meff = \ev_z \left[ \left( f(\bootedparam)-f(\fittedparam) \right)^T \matr{\Sigma}^{-1} \left( \bootdat-f(\fittedparam) \right) \right].
\end{equation}
%


\bsp	
\label{lastpage}
\end{document}